\documentclass[useAMS,usenatbib]{mn2e}
\usepackage{times}

\title[Morphology of low-redshift compact galaxy clusters]{Morphology of low-redshift 
compact galaxy clusters\\
{\Large I. Shapes and radial profiles}}

\usepackage{lscape}
\usepackage{epsfig}

\author[V. Strazzullo et al.]{V. Strazzullo$^{1}$\thanks{E-mail:
strazzul@na.astro.it (VS); paolillo@na.infn.it (MP)}, M.
Paolillo$^{1,2,3}$, G. Longo$^{1,3,4}$, E. Puddu$^{4}$, S. G. Djorgovski$^{5}$,
\newauthor R. R. De Carvalho$^{6}$, R. R. Gal$^{7}$\\
$^{1}$Dipartimento di Scienze Fisiche - Universit\`a Federico II, Polo delle Scienze e della Tecnologia, via Cinthia, I-80126 Napoli, Italy\\
$^{2}$Space Telescope Science Institute, 3700 San Martin Dr., Baltimore, MD 21218, USA\\
$^{3}$Istituto Nazionale di Fisica Nucleare (INFN) - Sezione di Napoli, Italy\\
$^{4}$Istituto Nazionale di Astrofisica (INAF) - Osservatorio Astronomico di Capodimonte, via Moiariello 16, Napoli, Italy\\
$^{5}$Palomar Observatory, Mail Stop 105-24, California Institute of Technology, Pasadena, CA 91125, USA\\
$^{6}$Instituto Nacional de Pesquisas Espaciais - DAS (CEA-NOVO), Av. dos Astronautas 1758, S.J.Campos SP 12227-010, Brazil\\
$^{7}$Department of Physics, U.C. - Davis, One Shields Ave., Davis, CA 95616, USA\\
}

\begin{document}

\date{Accepted . Received }

\pagerange{\pageref{firstpage}--\pageref{lastpage}} \pubyear{0000}

\maketitle
\label{firstpage}

\begin{abstract}

The morphology of clusters of galaxies may be described with a set of
parameters which contain information about the formation and
evolutionary history of these systems. In this paper we present a
preliminary study of the morphological parameters of a sample of 28
compact Abell clusters extracted from DPOSS data, measured with a
procedure based on the use of the {\small CIAO-Sherpa} software, developed at
the Center for Astrophysics (CfA) for X-ray data analysis. The
morphology of galaxy clusters is parameterized by their apparent
ellipticity, position angle of the major axis, centre coordinates,
core radius and $\beta$-model power law index. Our procedure provides
estimates of these parameters (and of the related uncertainties) by
simultaneously fitting them all, overcoming some of the difficulties
induced by sparse data and low number statistics typical of this kind
of analysis. The cluster parameters were fitted in a $6 \times 6$ Mpc$^{2}$ region, measuring the background in a $4 $ Mpc $< R  < 5$ Mpc annulus. We also explore the correlations between shape and
profile parameters and other cluster properties. Our results can be
summarized as follows: one third of this compact cluster sample has
core radii smaller than 100 kpc, i.e. near the limit that our data
allow us to resolve, possibly consistent with cusped models.  The
remaining clusters span a broad range of core radii up to $\sim 1500$
kpc, including some apparently regular clusters with well resolved
core radii.  More than 80 per cent of this sample has ellipticity
higher than 0.2. The alignment between the cluster and the major axis
of the dominant galaxy is confirmed at a high significance level,
while no correlation is observed with other bright cluster members. No
significant correlation is found between cluster richness and
ellipticity. Instead, cluster richness is found to correlate, albeit
with large scatter, with the cluster core radius. Finally, in contrast
with claims in previous works, a flat universe seems to be favoured,
and in any case is not excluded, by the power-law index $\beta$ of our
number density profiles.
\end{abstract}


\section{Introduction}

In CDM cosmological models, the formation of cosmic structures is
associated with that of dark matter halos, which form hierarchically
from the gravitational collapse of primordial density
fluctuations. Unlike galaxies, clusters are dark matter dominated
systems in which the baryons are not radially segregated. Also, unlike
galaxies, they form by dissipationless collapse. Thus, clusters in
principle provide a more direct probe of the primordial density
fluctuations.

The most accurate methods to derive the mass profile of galaxy
clusters (and also their internal structure) require either
strong/weak lensing analysis or, under the assumption of
thermodynamical equilibrium of the system, the velocity dispersion or
the X-ray gas density and temperature profiles. Apart from the
validity of the equilibrium assumption, the acquisition of the
necessary data is often an expensive task for large, statistically
significant samples of clusters.  Alternatively, under the assumption
that galaxies trace the underlying mass distribution, cluster-sized
dark halos density profiles can be derived from galaxy number
counts. While this method is not exempt from systematic uncertainties,
it relies on few assumptions and, unlike the other methods, can be
easily applied to large samples of clusters extracted from wide-field
surveys.

\citet{NFW96} showed via
extensive numerical simulations that in a CDM universe, halos of
different mass (from dwarf galaxies to rich clusters) follow a
universal cusped profile (hereafter NFW):

\begin{equation}
\frac{\rho}{\rho_{crit}}=\frac{\delta_{c}}{(r/r_{s})(1+r/r_{s})^{2}}.
\end{equation}

The NFW model and its universality have been challenged by several
authors, especially because of the rotation curves of low surface
brightness galaxies and dwarf galaxies, which show a relatively flat
density distribution \citep[see e.g.,][]{moore99, swaters2000,
blok02}.  As applied to galaxy clusters, even if the NFW model is
often accepted as a good fit to mass density profiles
\citep[e.g.][]{Cnoc97a, marke99, geller99, lewis03}, controversy still
exists and has been recently revived by several authors \citep[see for
instance][]{jordi95, shapiro00, wu01, sand2002, biviano03}.

From the number density and luminosity profiles of rich galaxy
clusters in the ENACS (ESO Nearby Abell Cluster Survey) catalog, Adami et al. (1998, 2001, hereafter AMKB98, AMUS01) found
that, if the faint cluster members are included, a core model of the
form:
\begin{equation}
s(r)=s_{0}\left(\frac{1}{1+(r/r_{core})^{2}}\right)^{\beta}
\label{betamodel}
\end{equation}
reproduces the galaxy number density profiles more accurately than
cusped (NFW-like) models; the trend turned out to be even more evident
for luminosity profiles. This result is in agreement with models
predicting different evolutionary scenarios for the luminous and the
faint populations of galaxies in clusters.  In fact, AMUS01 suggested
that galaxy number density and luminosity profiles could be originally
cusped, as expected from simulations of dark matter halos
\citep{NFW96,NFW97} and that, afterward, the cusp in the number
density profile could be erased by merging events.  This would produce
a cusp in the bright galaxy luminosity profile, as well as the core
observed in the faint galaxy luminosity profile through tidal
disruption of small galaxies near the cluster centre.

The cluster morphological parameters thus bear relevant information on
the cluster dynamical status and formation scenarios. For instance,
the core radius $r_{core}$ is a characteristic scale-length of the
galaxy distribution inside the cluster, and it has been shown to be
tightly correlated with the virial radius \citep[see][hereafter
G95]{girardi95}.  Other relevant information may be obtained from the
cluster shapes once they are approximated as spheroids.  Rich clusters
have been found to show ellipticities up to 0.8, with a mean value of
0.4, and a correlation between ellipticity and richness (the richer the cluster, the more 
spherical the shape -- e.g. \citealp{strubleftaclas94,PAM95}).
 Moreover, the cluster shape
distribution is directly connected to the cosmological parameters: a
low density universe produces more concentrated, spherically symmetric
clusters than a $\Omega = 1$ scenario \citep[e.g.][]{evrard93}.

In this work we present a preliminary study of the morphological
properties of a sample of 28 nearby compact Abell clusters extracted
from the Digitized Palomar Sky Survey (DPOSS) \citep{djorgovski}. The
advantages of using DPOSS data are mainly in the wide sky coverage
(which allows an extended region around each cluster to be studied),
the homogeneous photometric system of the whole data set, and good
control of the catalogue completeness and selection criteria
\citep{weir95,maurizio2001}.

We point out that we do not attempt to test which model best describes
the actual cluster profile. Instead, {\it we adopt the beta model as a
statistically adequate, analytically convenient description of the
galaxy distribution} and investigate the correlations between shape,
profile parameters and other cluster properties. The comparison
between different cluster profile models requires careful stacking of
individual clusters to obtain a sufficient S/N ratio (see for instance
\citealt{Cnoc97b,PAM99}), and is beyond the purpose of this
work.

The paper is structured as follows: in section $\S$ \ref{data} we describe the data
and the cluster sample, in section $\S$ \ref{radial_profiles} we describe the method used to
model cluster radial profiles and to derive morphological parameters,
while in sections $\S$ \ref{results} and $\S$ \ref{conclusions} we present our results and summarize our
conclusions. Throughout this paper we use $H_{0}=50$, $\Omega_{M}=1$, $\Omega_{\Lambda}=0$. We do not adopt the usual concordance cosmology in order
to simplify the comparison with Paolillo et al. (2001), on which this
work is partially based. Cosmological parameters only affect our
results via the conversion from angular to linear distances. Using a
H$_{0}$ = 70, $\Omega_{\Lambda}$ = 0.7, $\Omega_{M}$ = 0.3 cosmology,
distances would be smaller by approximately 20-25 \% in our redshift
range.

\section[]{Data and cluster sample selection}
\label{data}

This work is based on catalogues produced from DPOSS data
\citep{djorgovski} with the SKICAT package \citep{weir95}.  We started
with the initial sample of 80 Abell clusters described in
\citet{Piranomonte2001}, which had available calibration frames and at
least one reliable spectroscopic redshift \citep[see][]{maurizio2001}
at the beginning of this work. We note that a much larger cluster
sample extracted from DPOSS catalogs is currently available and its
analysis is postponed to a future work \citep{gal03}.

The photometric completeness limit was estimated for each cluster and
in each band independently; typical values are $m_{r} \simeq 20$ and
$m_{g} \simeq 20.5$ in the Gunn-Thuan photometric system
(\citealt{maurizio2001}; Paolillo et al., in preparation); K-corrections
were not taken into account since, due to the small redshift range
covered by the clusters in our sample ($z < 0.27$, with only 2
clusters at $z > 0.2$), they are negligible; furthermore, the lack of
morphological information for the individual galaxies would introduce
unnecessary ambiguities.

For each cluster we extracted from the SKICAT catalogues all objects
brighter than $m_{g}=20$ contained within a square region of 10
$\times$ 10 Mpc$^{2}$ centered on the cluster centre (as listed in the
Abell catalogue), detected in both the g and the r bands (the i-band
was not used since it is much shallower than the others) and
classified as galaxies in both filters.  The large region around each
cluster (10 $\times$ 10 Mpc$^{2}$) was selected in order to sample the
cluster galaxy distribution out to a significant distance from the
cluster centre (1 Abell radius = 3 Mpc with our cosmology), and to
have a control field around each cluster wide enough to achieve a
proper background determination. As discussed later ($\S$
\ref{mod_rad_prof}), the proper cluster fitting was performed on a
smaller region ($6\times 6$ Mpc).

Galaxy clusters may exhibit different degrees of internal
subclustering \citep{baier79,gellerbeers82,westetal88,dressler88,
bird94,pinkney96,solanes99,knebe2000}. In particular, cluster
substructure may generally be divided into four categories: dynamical
subclumps residing in generally relaxed systems; young clusters in an
early merging state; dynamically bound units in the cluster outskirts;
and chance projections of groups. Discriminating between these four
different cases is extremely difficult in the absence of velocity
information. Moreover, subclustering at large distances from the
cluster centre ($>1$ Mpc) may often be due to fluctuations in the
extended host supercluster \citep{westbothun}.

\begin{figure} 
\centering
  \begin{center}
 \epsfysize=8cm
\centerline{\epsfbox{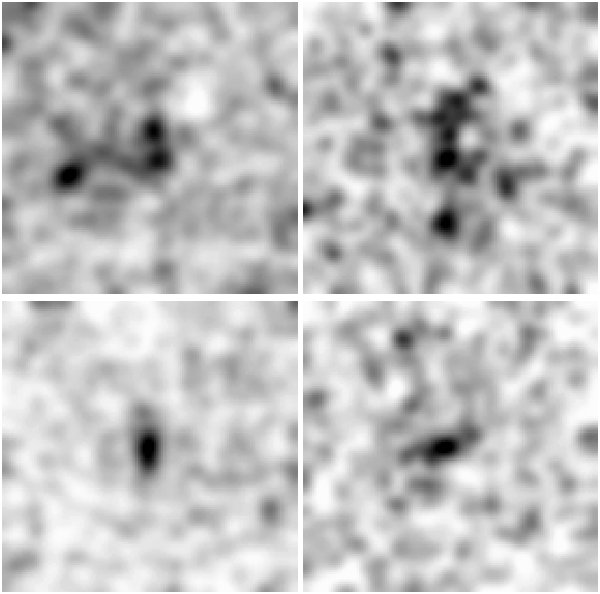}}
\caption{Typical examples of cluster galaxy number density maps, after smoothing with a Gaussian
filter with $\sigma$ = 250 kpc (in the cluster rest-frame). The
upper panels show two typical subclustered systems (A1035
and A1081) which are rejected by our selection criteria, while the lower panels present two compact clusters (A286 and
A1661) which are included in our sample. Each map covers an area of $10 \times 10$ Mpc$^{2}$.   \label{fig:subnosub}}
  \end{center}
\end{figure}

Obviously, the presence of substructures would make the modeling of
the radial profile almost meaningless, since there would be no radial
symmetry nor a well defined core radius associated with a central
relaxed region of the system. For the purpose of this work, we
excluded from our sample all clusters with obvious large substructures
by visually inspecting the galaxy surface density maps (smoothed with
a Gaussian filter with $\sigma$ = 250 kpc) and rejecting those with
multiple isolated peaks within 3 Mpc of the cluster center (see figure
\ref{fig:subnosub}).

While this is an empirical criterion, a more objective method is not
straightforward in our case, due to the low number of galaxies and
the lack of spectroscopic information.  Our criterion allows the
selection of clusters that are broadly regular and single peaked on
large scale, and therefore our sample includes several examples of
regular clusters like A1914, A1835, A566
\citep{buotetsai1996,majerowicz2002,zhou2003}. However, we also have
some systems which are not truly relaxed, generally known from X-ray
or dynamical analysis such as the disturbed morphology of A2061
\citep{postman1988}, the late merger A2142 \citep{buotetsai1996}, and
the bimodal cluster  A98 \citep{beers1982,krempec1995}.
We prefer to use a homogeneous criterion for the whole sample without
discarding individual clusters for which additional data are
available, even if this means that we have to accept some degree of
`imperfection', e.g. low density tails or some level of substructure
in the cluster core.

\begin{figure}    
\includegraphics[width=8cm]{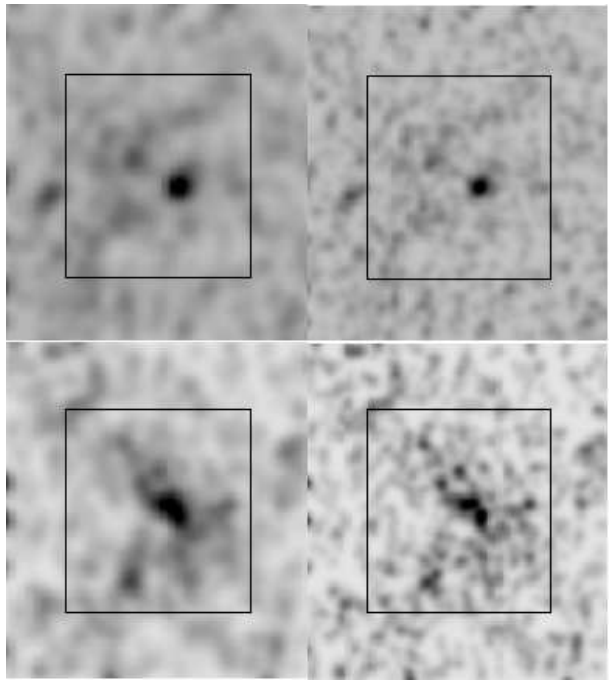}
\caption{Smoothed maps for A2083 (top) and A175 (bottom). Smoothing is gaussian with $\sigma$ = 250 kpc (left) and $\sigma$ = 150 kpc (right). The lower right panel shows cluster substructure on scales of order 150 kpc. The whole frame is $10 \times 10$Mpc$^{2}$ wide, while the inner square is $6 \times 6$Mpc$^{2}$. \label{fig:densitymaps}}

\end{figure}

It may be argued that a smaller smoothing scale may reveal smaller scale substructures. In figure \ref{fig:densitymaps} we show the comparison of density maps smoothed with 250 kpc and 150 kpc smoothing scales for two representative cases. We find that, using a 150 kpc scale, only five clusters (A175, A655, A910, A1661, A2177) reveal significant substructure that may in principle question their inclusion in the sample. Our data do not allow to study smaller scale structures in the cluster core since the distance between galaxies is on average greater than 50 kpc (i.e. on scales smaller than 150 kpc shot noise dominates the galaxy distribution).  Note however that a smaller smoothing window only affects the sample selection and not the actual fitting (see section $\S$ \ref{mod_rad_prof}).

The final sample of clusters which will be the subject of the
following analysis consists of 28 objects listed in table
\ref{tab:compactsample}.

\section[]{The determination of radial profiles}
\label{radial_profiles}
\subsection{Model and statistics}
\label{mod_rad_prof}

Modeling the 2D distribution of galaxies in clusters is a problematic
task due to the small number statistics which affects these sparse
data. The classical approach which obtains the number count radial
profile by binning and counting galaxies in concentric circular
shells, results in the loss of relevant morphological
information. Moreover, because of contamination from
background/foreground objects, there is a significant decrease in the
S/N ratio of the resulting profiles.

Maximum likelihood methods as initially proposed by Sarazin (1980),
applied directly to the galaxy positions without any binning, allow
these problems to be partially overcome. In general, these methods
rely on three fundamental assumptions: i) the observed positions of
galaxies are statistically independent (i.e. no substructure is
present); ii) the local background is assumed to be uniform; iii) the
shape of the galaxy number density function is assumed to be known and
characterized by a small number of parameters.

In the literature, radial profiles of galaxy clusters have mainly been
determined by means of different implementations of this
method. However, even if in principle it would be possible to fit all
the parameters simultaneously, in practice the minimization often has to be
accomplished insteps, fitting the shape (centre coordinates,
ellipticity and position angle) and the profile (core radius, power
law index etc.) parameters separately (e.g. AMKB98; see also G95,
AMUS01 and \citealt{girardi01}). Furthermore, in such cases the strong
correlation between the model parameters may produce spurious results.

Taking advantage of the formal analogies existing between X-ray data
(photon hit positions and photon energies) and galaxy catalogues
(galaxy positions and magnitudes) in terms of sparse distribution and
low statistics, we tailored to our specific needs the Sherpa package
present in the {\small CIAO}\footnote{We used CIAO version 2.2.1. Full
documentation can be found at http://cxc.harvard.edu/ciao/} software,
developed at the Chandra X-ray Center. Our procedure allows us to fit
all the parameters of the galaxy distribution at the same time,
instead of requiring a multi-step approach.

The Sherpa software works on binned data. To minimize the loss of
information, we binned the galaxy positions using a square grid of bin
size 50 kpc, in order to have at most one galaxy per bin, except for a
few clusters in which the highest density bin may contain two
galaxies. This is in principle equivalent to using unbinned data.

We assume that galaxy clusters can be described by a standard two
dimensional $\beta$-model of the form:

\begin{equation}
f(r)=f(x,y)=\frac{\Sigma_{0}}{\left(1+(r/r_{core})^{2}\right)^{\beta}
} \label{eq:sherpamodel1}
\end{equation}

\noindent where:
\noindent
\begin{eqnarray}
  \label{eq:sherpamodel2}
  r(x,y)=\frac{\sqrt{\bar{x}^{2} (1-\epsilon)^{2} + \bar{y}^{2}  }   }{1-\epsilon}\\
\bar{x}=(x-x_{0})\cos\theta + (y-y_{0})\sin\theta\\
\bar{y}=(y-y_{0})\cos\theta - (x-x_{0})\sin\theta.
\end{eqnarray}

This implies that the characterization of the galaxy profile of each
cluster requires the simultaneous fit of the following parameters: \\
i) shape parameters: centre coordinates ($x_{0},y_{0}$), ellipticity
$\epsilon$, and position angle $\theta$; \\ ii) profile parameters:
core radius $r_{core}$, power law index $\beta$, central density
$\Sigma_{0}$ at $x_{0},y_{0}$), and local background density
$\Sigma_{bkg}$ (assumed to be uniform )\footnote{The external
constraint due to the normalization of the model to the observed total
number of galaxies, reduces the degrees of freedom of the
problem. Therefore, only three of the four profile parameters are
actually independent.}.\\

\begin{figure} 
  \begin{center}
 \epsfxsize=8cm
\centerline{\epsfbox{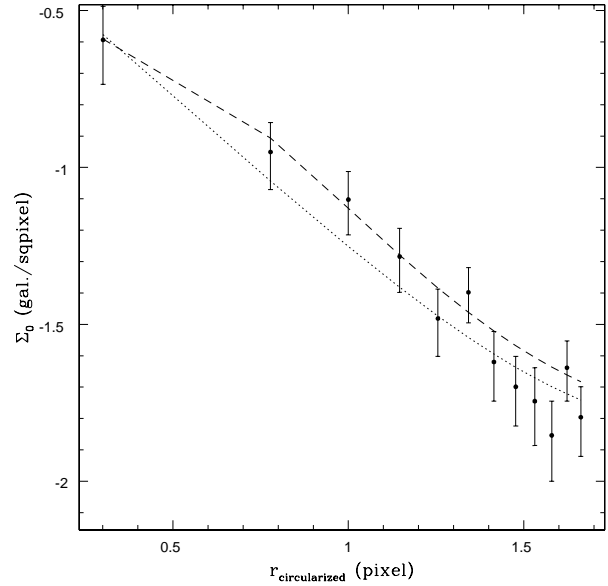}}
\caption{Example of a unidimensional fit on binned data for one cluster of the sample (A286), with both a cusped profile 
(dotted line) and a $\beta$-model (dashed line).     \label{fig:corecusp}}
  \end{center}
\end{figure}

As already explained in the introduction, our goal is not to determine
which function describes best the profile of galaxy clusters.  We are
heavily limited by the small number of galaxies and by the sparse
nature of the galaxy distribution, which makes it difficult to
discriminate between different profiles for most clusters in our
sample. As an example, in figure \ref{fig:corecusp} we compare the
best-fit NFW and $\beta$-model profiles, derived through an
azimuthally averaged unidimensional fit on binned data, for A286. Even
with this simplistic procedure, which minimizes the number of fitted
parameters, the two models are almost indistinguishable. In $\S$
\ref{results} we will discuss further the possibility that our
clusters follow a cusped distribution.

The use of parametric methods has also been criticized, in particular
when not dealing with truly sparse data or theoretically well-
established functional forms. This criticism arises mainly from the
fact that projection effects can transform two intrinsically different
spatial profiles to very similar projected profiles, and by the
application of parametric methods to a typical case of an
ill-conditioned problem such as the radial profile estimation
\citep{merritttremblay94}. Although in our case we are actually
dealing with sparse data, so that a non-parametric approach would be
quite difficult, we are aware that we are simply using a statistically
convenient empirical formula in a consistent manner.

In performing our fit we used a maximum likelihood approach, where the
likelihood function is the product of the individual probabilities
$P_{i}$ computed for each bin $i$ assuming that galaxy counts are
sampled from a Poisson distribution:

\begin{equation}
\mathcal{L}=\prod_{i} \frac{P_{i}^{N_{i}}} {N_{i}!} exp(-P_{i}).
\label{eq:Lcash}
\end{equation}

In equation (\ref{eq:Lcash}) $P_{i}$ is the sum of cluster and
background model amplitudes, and $N_{i}$ is the number of observed
galaxy counts in bin $i$. The Cash statistic \citep{cash}:

\begin{equation}
\mathcal{C}=2\left[\sum_{i}P_{i} - N_{i,S}lnP_{i}   \right]
\label{eq:Ccash}
\end{equation}

\noindent is derived from the likelihood function by taking the
logarithm of $-\mathcal{L}$ and dropping the factorial term, which is
constant in fits to the same dataset. Unlike the more traditional
$\Delta \chi ^{2}$, the Cash statistic may be used regardless of the
number of counts in each bin (for details, see the
\citealt{Sherpamanual}).  We note that, with poorly sample data, the
Cash statistic works better than a $\chi ^{2}$ statistic, even if
the latter is tailored to work with low number counts.

\subsection{Fitting procedure and validation}
\label{fit_val}

The application of the {\small CIAO-Sherpa} software for the determination of
cluster radial profiles was extensively tested on real and simulated
data. A first problem to solve was how to estimate the galaxy
background. Tests performed on the $10\times 10$ Mpc$^{2}$ region
(i.e. $R \la 5$ Mpc) showed that, for the majority of clusters, the fit was able to recover within 10$\%$ the estimate of the background level as obtained manually, by measuring the average galaxy
counts in an external annulus ($4$ Mpc $< R < 5$ Mpc) around the
cluster. Both the manual and fitted estimate are reported in table \ref{tab:results2}.

\begin{figure} 
\epsfxsize=10cm
\centerline{\epsfbox{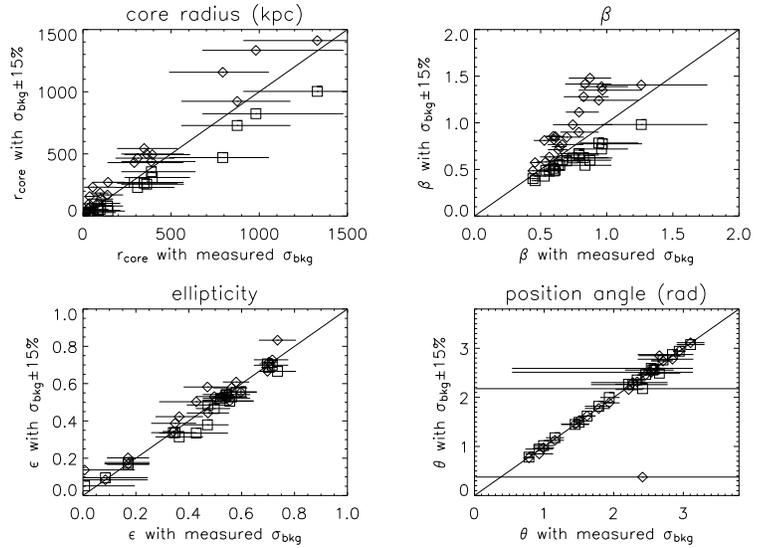}}
\caption{Effect of a $15\%$ background misestimate on the parameters: core radius (top left), $\beta$ (top right), ellipticity (bottom left) and position angle (bottom right). In all the panels, the diamonds mark the parameter value fitted with a background density overestimated by $15\%$ of its measured value vs. the usual measured value, while the squares mark the parameter value fitted with a underestimated background vs the usual value. All the fits are performed in the $6 \times 6 Mpc^{2}$ region. The solid line traces the bisector. \label{fig:bkg15}}
\end{figure}

For low S/N objects, while still retrieving the correct background, problems were encountered in fitting the other cluster parameters in the whole $10\times 10$ Mpc$^{2}$ region. This is a natural consequence of the fact that with low S/N objects, the background fluctuations undermine the cluster
detection itself. Therefore, in order to be consistent, we decided to
fix for all the clusters the background density to the value measured
in the $4 < R < 5$ Mpc annulus, and to fit the cluster in a smaller
region of $6 \times 6$ Mpc$^{2}$.
We tested the effect of a background under- or over-estimation by 15$\%$ ($\simeq 2\sigma$ upper limit on the background fluctuations measured in our catalogs over a $6 \times 6$ Mpc$^{2}$ window) of its measured value on the fitted parameters $r_{core}$, $\beta$, ellipticity and position angle: as expected, the ellipticity and position angle are scarcely affected, while there is a  systematic effect on $\beta$ and, to a lesser extent, on the core radius (see figure \ref{fig:bkg15}).

\begin{figure} 
\includegraphics[clip,width=9cm,bb=73 368 542 720]{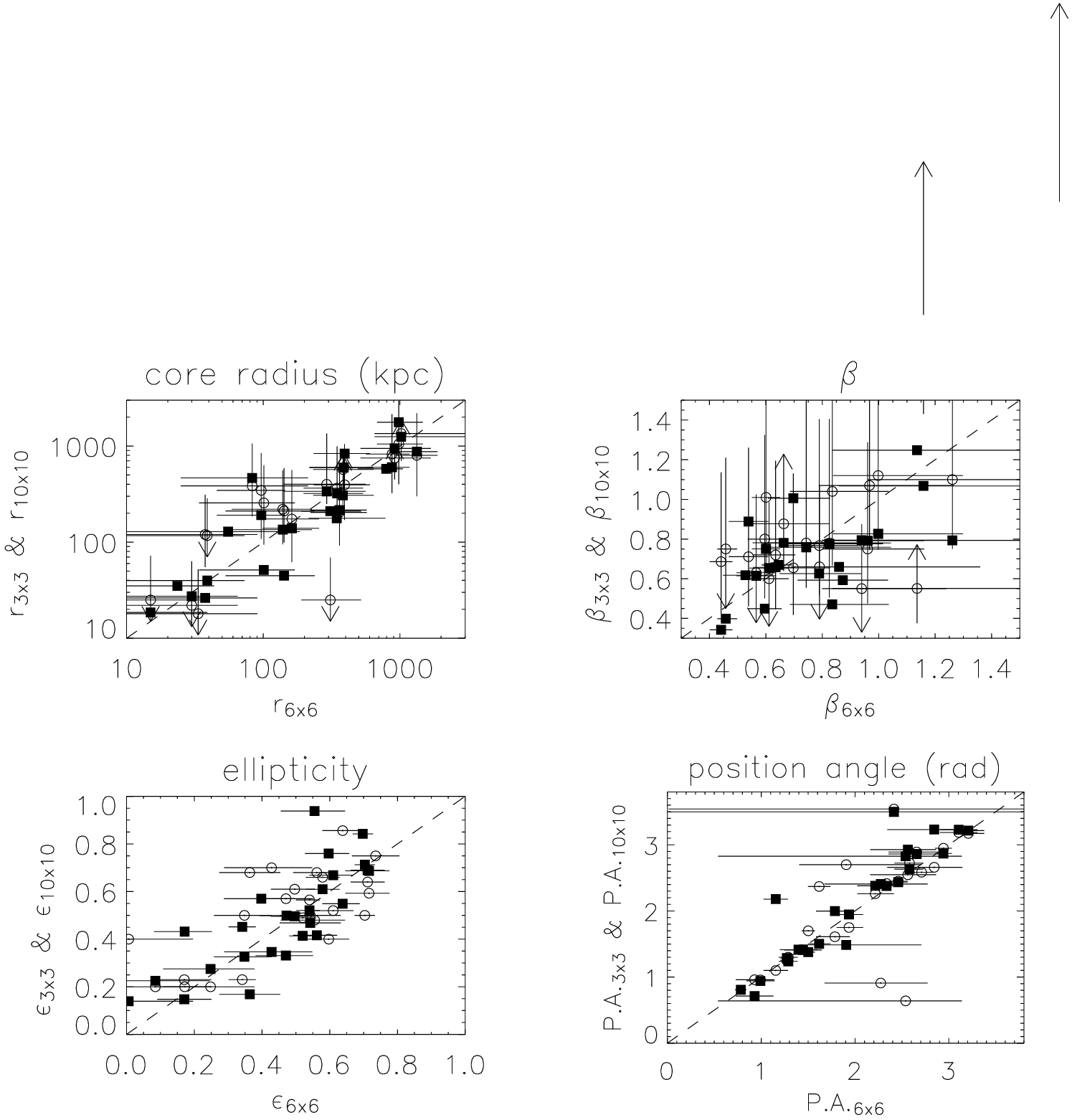}
\caption{Dependence of the parameters on the fitted region: core radius (top left), $\beta$ (top right), ellipticity (bottom left) and position angle (bottom right). In all the panels, the empty/filled symbols mark the parameter value fitted in the $3 \times 3 Mpc^{2}$ / $10 \times 10 Mpc^{2}$ regions respectively vs. the value in the $6 \times 6 Mpc^{2}$ region. The dashed line traces the bisector. \label{fig:regdep}}
\end{figure}

The $6 \times 6$ Mpc$^{2}$ was adopted as an optimal compromise
between the requirement to sample a relevant part of the cluster,
possibly out to the radius where the galaxy density approaches the
background level, and the attempt to maximize the S/N ratio of our
galaxy counts. Note that this region corresponds to a radius $R=3$
Mpc, i.e. one Abell radius in our adopted cosmology, and approximately
to the median cluster virial radius \citep{Cnoc97b}.  Furthermore this
choice helps to prevent strong background fluctuations or nearby
($3<R<5$ Mpc) overdensities from affecting the cluster parameter
evaluation.  The $4$ Mpc $<$ R $<$ $5$ Mpc annulus represents a region
distant enough not to be contaminated by the cluster itself while
sufficiently nearby to be representative of the local cluster
background.  
Since the use of a $6\times 6$ Mpc region may include small local
substructures (either due to the cluster or the background) which are
not discarded by our sample selection criteria, we performed an
additional fit on a smaller $3 \times 3$ Mpc$^{2}$ area. The results,
which are discussed in section $\S$ \ref{results}, show that the fits performed on the
smaller region, while severely affected by the lower statistics, are
basically consistent with those performed over the $6\times 6$ Mpc$^{2}$
area.
We show the dependence of the retrieved parameters on the size of the fitted region in figure \ref{fig:regdep}.
The estimates in the three different regions appear consistent within the errors in the vast majority of cases. The $\beta$ parameter shows indeed a large scatter, as this is the less stable of the fitted parameters, due to low S/N and incomplete profile sampling in small apertures.

The fit results do not depend on the bin size,
provided that the bin is small enough to avoid any loss of information
(i.e. $\sim 1$ galaxy bin$^{-1}$). Furthermore, the shape
parameters (centre coordinates, $\epsilon$ and $\theta$) derived from
the fit are consistent within the errors with respect to profile ($r_{core}, \beta$)
parameter variations (see figure \ref{fig:regdep}). 

We don't find any significant correlation of the fit result on the starting values: using different values within the permitted range (i.e. [0:3000] kpc for r$_{core}$, [0:4] for $\beta$, [0:1] for $\epsilon$, and [0:2$\pi$] for $\theta$), the difference in the retrieved output is 5-20\% for r$_{core}$, 1-15\% for $\beta$, less than 2\% for $\epsilon$, and less than 0.1\% for $\theta$ (the larger differences quoted for r$_{core}$ and $\beta$ occourring for the clusters with very large core radius).

Finally, we tested our procedure on a sample of 230 mock clusters.
Mock clusters are generated following a $\beta$-model profile, and all clusters are produced with
centre coordinates in the centre of the mock catalog region, which in pixel units is  ($x_{0},y_{0}$) = (60,60), ellipticity = 0.4 and $\beta$ = 0.8, representative of typical values in
the real sample, and position angle = 1. The core radius, central density and background density are randomly chosen in such a way that the total number of objects, the
background density $\sigma_{bkg}$, the cluster core radius, and the
S/N reflect those of the real clusters.  The fit on the mock catalogs
is performed exactly in the same way as for real clusters: background
density is fixed and all the other parameters are left free to float.

\begin{figure} 
\epsfxsize=8cm
\centerline{\epsfbox{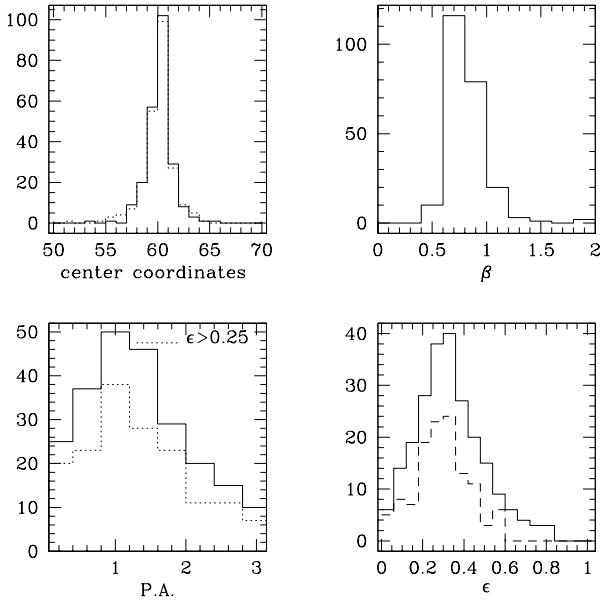}}
 \caption{Distribution of the mock cluster parameters. Upper panels:  distributions of the retrieved $\beta$ values (right) and centre coordinates (left - solid and dotted line for $x_{0}$ and $y_{0}$, input value was (60,60)). Lower right panel: distribution of retrieved ellipticity; the solid line shows the whole sample, dashed line denotes the high S/N clusters. Lower left panel: distribution of position angles, solid line is for the whole sample, dotted line is for clusters with retrieved ellipticity greater than 0.25. \label{fig:mock1}}
 \end{figure}

\begin{figure} 
\epsfxsize=8cm
\centerline{\epsfbox{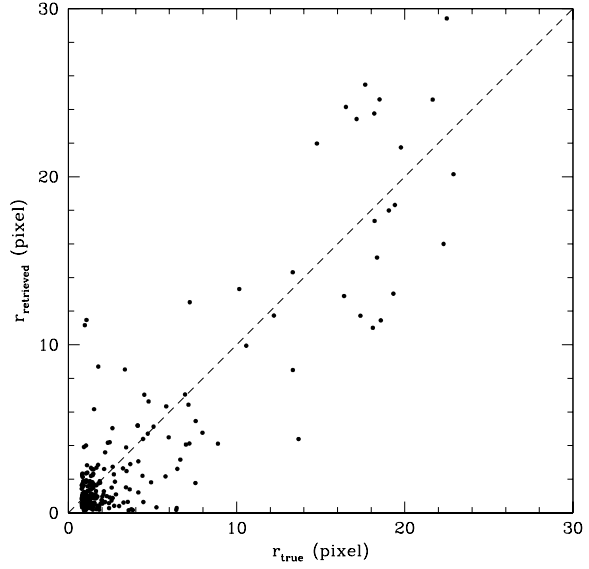}}
 \caption{Retrieved vs. true core radius for the mock cluster sample. The dashed line traces the bisector. \label{fig:mock2}}
 \end{figure}

The mean ellipticity retrieved is $0.33 \pm 0.16$, the mean $\beta$ is
$0.83 \pm 0.26$, the mean centre coordinates are $(60.2,60.1) \pm
(1.2,1.5)$, the mean position angle is $1.3 \pm 0.7$. The distributions of the retrieved ellipticity, $\beta$,
centre coordinates and position angle are shown in figure \ref{fig:mock1}, and are to be compared with the above reported true-values of  ($x_{0},y_{0}$) = (60,60), ellipticity = 0.4, position angle = 1., and $\beta$ = 0.8. In figure \ref{fig:mock2} we show the retrieved vs. true core radius.
These plots show that our fitting procedure yields reliable results
without significant systematic effects, except possibly for the
ellipticity which is slightly skewed toward lower values. This is a
result of the background contribution which tends to azimuthally
smooth the galaxy distribution. This effect however does not affect
our results, as discussed in $\S$ \ref{results}. The large dispersion
in the P.A. distribution is mainly due to low S/N. In fact, additional
simulations with very high S/N mock clusters show that for $\epsilon >
0.3$ the position angle is correctly retrieved within 10 degrees at
most.  Note that while the core radius exhibits a large relative
dispersion at small radii, due to the difficulty of obtaining a
reliable estimate of $r_{core}$ for small core radii (due to the low
S/N within $r_{core}$), the retrieved value is in most cases
representative of the true value, and the scatter found for mock
clusters is consistent with the average measurement errors that we derive
for the real sample.

\subsection{Fitting results}
\label{fit_res}
Summarizing the discussion in section $\S$ \ref{fit_val}, we performed the fit for
all clusters in a $6 \times 6$ Mpc$^{2}$ square region, with a bin
size of 50 kpc and with the background density fixed to the value
measured for each cluster in an external annulus (4 Mpc $< R <$ 5
Mpc). Note that the use of the same physical area ($6 \times 6$
Mpc$^{2}$) for all clusters implies that we are considering a
different density threshold for each cluster. However, this does not
affect our results since we sample the cluster out to a radius where
the S/N is negligible, while the use of a fixed density threshold
would introduce other effects, such as a dependence on the background
fluctuations.

The resulting parameters are listed in table \ref{tab:results1}.
Error estimates are calculated by leaving only the highly correlated
parameters free to vary (i.e. $\epsilon - \theta$ and $r_{core} -
\beta - \Sigma_{0}$), and fixing all the others. Even if these are not
the formally correct errors, they provide a meaningful uncertainty on
the parameter estimates. The formal errors (where all the fit
parameters are left free to vary) would often be unconstrained, given
the low counts and the large number of parameters being fit.

\begin{figure*} 
\epsfxsize=8cm
\centerline{\epsfbox{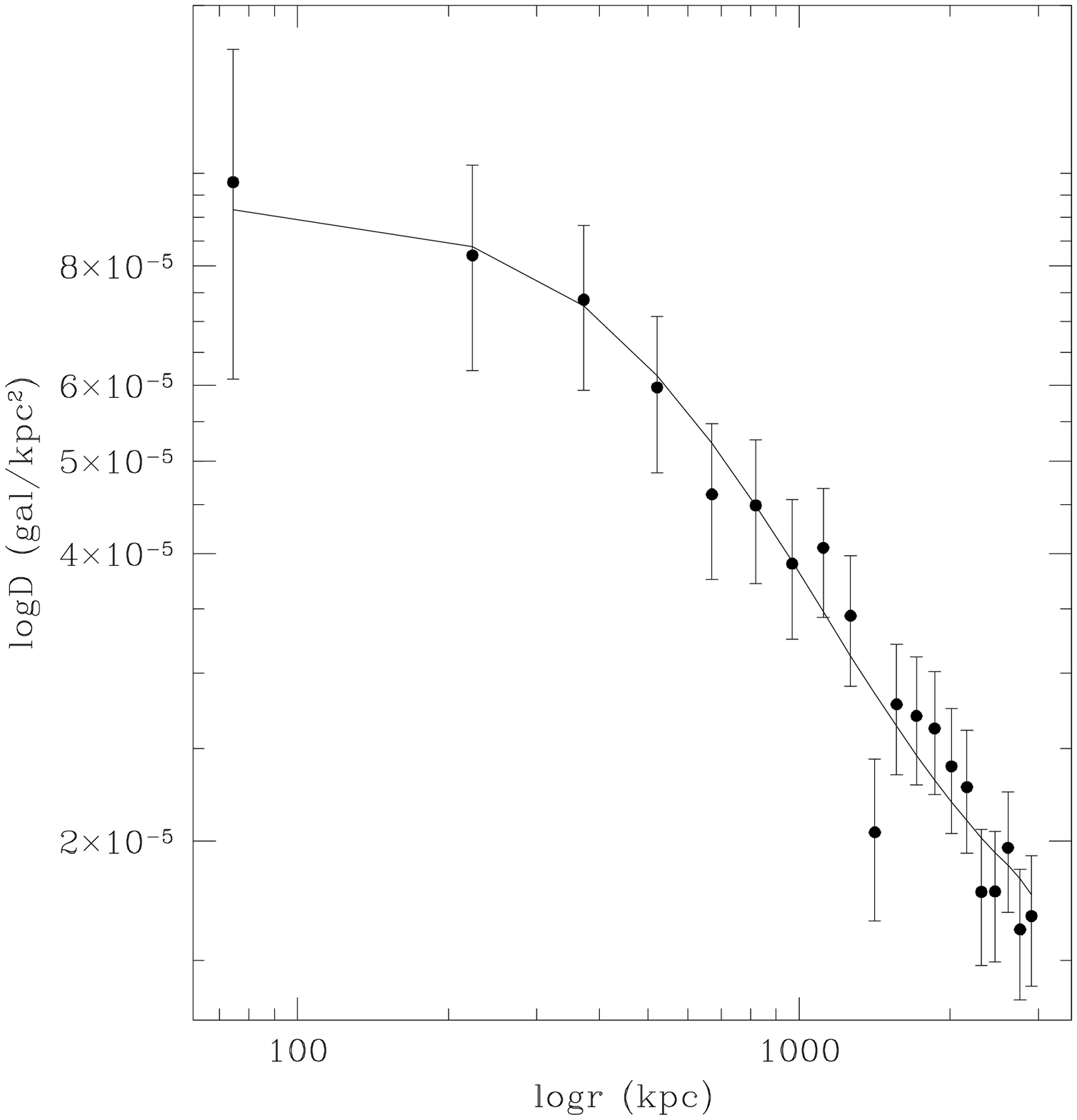}%
\epsfxsize=8cm
\epsfbox{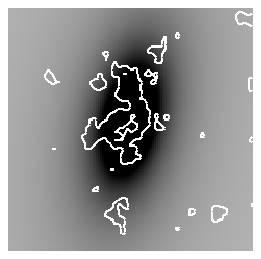}
}
\epsfxsize=16cm
\centerline{\epsfbox{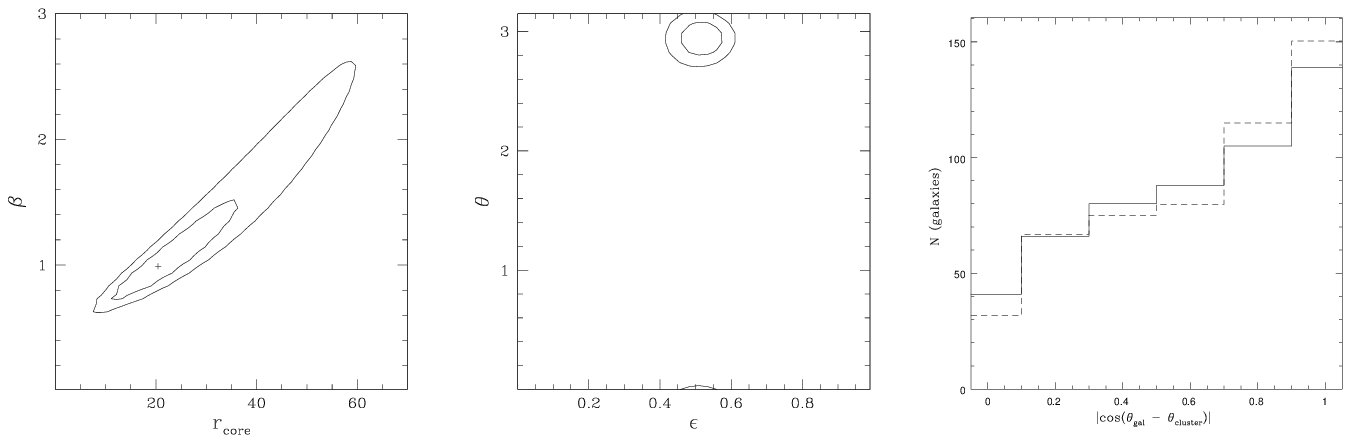}}
\caption{Cluster A98. Top left panel: radial profile (the solid line
marks the model while dots mark the data, see text for details). Top
right panel: cluster isocontours in the fitted region ($6 \times 6$
Mpc$^{2}$) plotted on the model. Bottom left and central panels: 1 and
2 $\sigma$ confidence regions for the couples ($r_{core} - \beta$)
(with $\Sigma_{0}$ free to vary) and ($\epsilon - \theta$)
respectively (r$_{core}$ is in units of bins (1 bin = 0.325 arcmin
$\simeq$ 50 kpc), while $\theta$ is expressed in radians).  Bottom
right panel: the misalignments of galaxy P.A. with cluster
P.A. (solid line) compared with those expected for a random
distribution of galaxy position angles (dashed line).  \label{fig:A98}}

\end{figure*}

As a template case, we show in figure \ref{fig:A98} the results of the
fitting procedure for the cluster A98. In the upper left panel, the
solid line traces the best-fit profile and the dots show the measured
density profile. The data and the model are binned using the same set
of annuli centered on the cluster centre defined by the fit. We point
out that this binning is only used for visualization purposes and it
is not involved in the determination of the best-fit parameters.  In
the upper right panel, we plot the cluster isodensity contours against
the model.

In the lower left and middle panels, we plot the confidence contours
at $ 1 \sigma$ and $2 \sigma$ for the highly correlated parameters,
namely $\epsilon - \theta$ and $r_{core} - \beta$ (with $\Sigma_{0}$
free to vary).  Finally, in the lower right panel we plot the
distribution of $|\cos(\theta_{galaxies}-\theta_{cluster})|$ (solid
line) and the expected histogram for a random distribution of the
galaxy position angles (dashed line).  For the other clusters in the
sample we show only the three most significant plots, namely the
cluster isodensity contours against the model, the fitted profile, and
the $|\cos(\theta_{galaxies} - \theta_{cluster})|$ distribution
(fig. \ref{fig:indivcluster1}).

\section[]{Discussion}
\label{results}
In figure \ref{fig:histopar} we present the distribution of
ellipticities (left panel), slopes $\beta$ (middle panel), and core
radii (right panel) derived from our fits.  Even though the sample was
selected to include clusters with an overall regular morphology, we find a wide range
of ellipticities, with mean value $\overline{\epsilon}=0.47$ and
dispersion $\sigma_\epsilon=0.2$; note that, given the systematic
error measured by our simulations ($\S$ \ref{fit_val}), the real
ellipticity distribution is expected to be slightly skewed toward
larger elongations by $\Delta\epsilon\sim 0.07$. While this result is
in fair agreement with the $\epsilon$ distribution measured by
\citet{PAM95}, it is in contrast with AMKB98, who found on average
low ellipticities, consistent with $\epsilon\simeq 0$. However, AMKB98 and AMUS01 were mostly interested in the central cluster region, in order to compare cusped and core profiles; as
these authors point out, their study is limited to the very central region of the cluster ($\simeq 500$
kpc) specifically chosen to avoid substructures. This implies that, apart from possible aperture biases which could affect the result, the cluster region that they fit has intrinsically very small ellipticity. 

In fact, in our sample six out of the eight less regular clusters mentioned in $\S$ 2 have ellipticity above the mean.

As noted in section $\S$ \ref{fit_val}, in order to better evaluate possible effects
introduced by overlooked background substructures in the cluster
outskirts, we performed an additional fit over a smaller $3 \times 3$
Mpc$^{2}$ region. While the parameters derived in this way are more
uncertain because of the lower statistics, and possibly biased by the
exclusion of the cluster external region, the ellipticity distribution
and the presence of the peak at $\epsilon \simeq 0.5$ are confirmed
also using the parameters derived from this smaller region (the dashed
line in figure \ref{fig:histopar}).  Furthermore, in figure
\ref{fig:pa1pa3} we plot the position angles $\theta$ derived within
the $ 6 \times 6 $ Mpc$^{2}$ region against those derived within the $
3 \times 3 $ Mpc$^{2}$ region (clusters A2061, A2083, A2178 and A910
are not included because no reliable parameters could be determined
for them in the $3 \times 3$ Mpc$^{2}$ region). The two estimates are
fully consistent and the few points scattered above and below the line
are clusters with $\epsilon \le 0.25$, i.e. objects for which the
error on the position angle is systematically larger (the higher the
ellipticity, the better constrained the position angle of the major
axis).  This result shows that there is no significant substructure in
the cluster outskirts which would affect the cluster P.A., even
though the contribution of cluster galaxies is non-negligible between
1.5 and 3 Mpc from the cluster centre: a significant fraction (between
20 and 60 per cent, with a median value of approximately 50 per cent)
of cluster members is found in this external region. Therefore, any
study of cluster structure sampling less than 1.5 h$^{-1}$ Mpc in
radius is missing a significant cluster contribution.

The power-law slope $\beta$ has a peak at $\sim 0.7$, with
$\overline{\beta}=0.8$ and dispersion $\sigma_\beta=0.27$; both the
average value and the shape of the distribution are in very good
agreement with the results obtained by \citet{ppopesso} based on the
ROSAT-SDSS galaxy cluster sample, as well as with the results of
\citet{girardi01} based on a sample of moderate/high redshift clusters
.  On the other hand, our results are again in contrast with those of
AMKB98, who found $\beta=1.02\pm 0.08$ for their sample of 60
clusters. Even though our distribution is quite broad, the average
value of $\overline{\beta}$ is inconsistent at the $>3\sigma$ level\footnote{The error on $\overline{\beta}$ is $\sigma_{\beta}/\sqrt{N}$ where $N$ is the number of clusters in the sample. }
and none of the clusters in the AMKB98 sample has $\beta <0.84$.
While our catalogues are somewhat deeper ($\sim 0.5$ mag) than the
ones used by AMKB98, the main differences between us and AMKB98 are
that the latter sample a smaller region, thus excluding the
contribution of the cluster outskirts, and by the different approach
used to estimate the background. Note however, that the fits performed
on our smaller $3\times 3$ Mpc region, which is closer to the AMKB98
area ($r>5 r_{core}\sim 1000$ kpc\footnote{We rescaled the AMKB98
sizes by a factor 2 since they assume a cosmology with $H_0=100,
q_0=0$}), still yield $\overline{\beta}=0.75$. The two $\beta$
estimates for the individual clusters are consistent within the errors
and there is no evident correlation of $\beta$ with the size of
the sampled region.  Note that, according to figure 10 of AMKB98 (based
on simulations by \citealt{Crone94,jing95,NFW95,walter96}) our $\beta$
distribution is compatible with flat universe models either with
zero (standard CDM) or non-zero cosmological constant ($\Omega_{m}=0.2
\div 0.3$ and $\Lambda =0.7 \div 0.8$), in contrast to the AMKB98
result which favoured an open cosmology. We remind, however, that the validity of cosmological constrains
obtained through this method is still debated, and depends on the used
simulations.

The median core radius is $r_{core}=310$ kpc, but the distribution is
strongly skewed toward low values with a marked peak at $r_{core}\sim
100$ kpc. Approximately 1/3 of our sample has $r_{core}< 100$ kpc,
corresponding to 2 pixels in our fitted maps; for such clusters
$r_{core}$ is thus only marginally resolved, as discussed in $\S$ 3.1,
so that we cannot exclude a cusped profile.  On the other hand 7
clusters have core radii greater than 500 kpc; among these we find six of the less regular systems discussed in $\S$ 2. 

Some correlation between the dynamical state of the cluster and its core radius is clearly expected, and in fact the most regular clusters (i.e. A79, A763, A971, A1677, A2065, A2083, A2223) all have very small core radius. The intermediate values of core radii refer to clusters which are generally single peaked, but exhibit some broader asymmetric overdensity region, or a filamentary structure. These systems also include the examples of known regular clusters mentioned in $\S$ 2.

Our core radii, including the whole sample,  are on average larger than those measured by
AMKB98 and AMUS01. However, our mean core radius excluding the eight less regular systems would be 213 kpc, with a median of 140 kpc. Nonetheless, AMKB98 found only 6 clusters out of 60 with $r_{core}<100$ kpc. Thus our sample spans a broader range of core radii (mostly due to the inclusion of less regular systems) but is also more peaked toward low values. 
This could be due to intrinsic differences in the nature of the cluster sample (indeed \citealt{solanes99} found a very low level of substructure in the ENACS clusters), as well as to the different area used, and possibly to differences in the method adopted, as the intrinsic coupling
between $r_{core}$ and $\beta$ in the model (we show for reference the $r_{core}$ vs. $\beta$ relation in figure \ref{fig:corevsbeta}) is expected to produce
larger core radii in correspondence of the steeper power-law slopes.  

On the other hand, the peak of the core radii distribution at 100 kpc is consistent with the small
values ($<100 h^{-1}$ kpc) of the core radius recently reported by
several authors (e.g.  \citealt{girardi01,katgert04}). Our tests (cfr. figure \ref{fig:bkg15}) show that a systematic underestimate of our background level by $\simeq 15\%$ (or viceversa an overestimate of AMKB98) would result in an average $r_{core}$ consistent with the AMKB98 value, provided that the most irregular clusters discussed in $\S$ 2 are excluded from the sample. However the $\beta$ distribution remains inconsistent at the 3$\sigma$ level.  

The fits performed on the $3\times 3$ Mpc region are generally
consistent with those performed on the whole $6\times 6$ Mpc area,
even though somewhat smaller, yielding a median $r_{core}=220$
kpc. However a KS test still rejects the hypothesis that our $3\times
3$ Mpc sample and the AMKB98 one are drawn from the same distribution
at the 97\% level.  Note that a misplacement of the cluster center
would flatten the central density profile resulting in larger core
radii (see for instance \citealt{beersetonry86}). According to AMKB98,
displacements smaller than 100 kpc have very small effect on the
central profile even in the extreme case of a cusped
distribution. Thus while such an effect could explain the few clusters
with very large $r_{core}$, it is unlikely to affect the majority of
the clusters.

Significant alignment between clusters and their dominant galaxies has
often been observed \citep[see][and references
therein]{lambas88}. This effect seems to be independent of cluster
richness \citep{toddwest99} and is confirmed up to high redshift. It
involves only the cluster dominant galaxy, while the other bright
galaxies appear to be randomly oriented, so that the alignment has
been claimed to be produced by the same processes that created the
dominant galaxy \citep{kim2001}.  In figure \ref{fig:bcm} we show the
misalignment between the position angles of the clusters and those of
their brightest cluster members (BCM), for those clusters with
ellipticity large enough for the position angle to be well
constrained.  We note that the BCM definition itself depends on the
observed area and that it may happen that the BCM lies far from the
cluster centre, thus questioning the association of the galaxy with
the cluster (if no redshift measurement is available).  Because of
this reason, when the BCM was located in the cluster outskirts we used
the position angle of the giant elliptical closer to the actual
cluster centre, as derived from the fitting procedure. In our case,
however, this correction affects only slightly the alignment results,
since it involves less than one third of the `high ellipticity sample
(i.e. $\epsilon > 0.25$).  As can be seen, the correlation existing
between the two quantities is confirmed: the random distribution is
rejected at a significance level higher than 99.9 \%.

For the other galaxies in the cluster, inspection of the right panels
in figure \ref{fig:indivcluster1} shows that the distribution of their
position angles relative to the cluster orientation are consistent
with a random distribution, with the possible exception of clusters
A2065, A2142, A2178 and A2223. In fact, according to the
Kolmogorov-Smirnov confidence levels listed in table
\ref{tab:results2} (column 9), these four clusters show a position
angle distribution significantly different from the random case.

To compare the morphological parameters with the cluster richness we
considered two different estimates. The first ($R_{A}$) is
calculated using a criterion similar to the Abell definition,
counting how many galaxies fall in the cluster area, with magnitudes in
the range [$m_{3}$, $m_{3}+2$], and then subtracting the number of
field galaxies in the same magnitude range and in an equally large
region.  The Abell richness estimate, while being historically and
operationally correct, is known to suffer from several drawbacks,
mostly due to the use of a fixed apparent magnitude range \citep[see,
for instance][]{gal03} .

For these reasons we also use a second richness estimate $R_{LF}$ based
on the cluster luminosity function, which allows us to exploit the entire
galaxy sample down to the completeness limit. We used the luminosity
functions (LFs) derived by Paolillo et al. (2001; Paolillo et al., in
preparation) to compute, for each cluster, the ratio of the number of
galaxies brighter than its absolute completeness limit to the
number of galaxies brighter than the same limit in the deepest sampled
cluster \citep{garilli99}.  Assuming that the luminosity function has
a universal shape, this ratio provides an estimate of the richness of
each cluster relative to the richness of the deepest sampled one.
While $R_{LF}$ is not exempt from uncertainties, in particular due to
the assumption of the universality of the luminosity function (which
is challenged by recent data, e.g. \citealt{Piranomonte2001}, Paolillo
et al. in preparation), we note that this approach is equivalent to
the use of a fixed absolute magnitude range, and in fact $R_{LF}$ is
well correlated to the richness estimate based on $M^{\star}$ employed
by \citet{gal03}.  A comparison between $R_A$ and $R_{LF}$ is shown in
Figure \ref{fig:rich}.

In Figure \ref{fig:r0rich} we show the dependence of the cluster core
radius on the cluster richness. While the core radius shows little
dependence on the Abell richness, a mild correlation (correlation
coefficient of $0.62$) is found using the LF richness estimate. The
stronger correlation found with $R_{LF}$ could be due to the fact that
the latter quantity is a better estimator of the cluster richness
since it takes into account the entire detected galaxy population,
i.e. the same population used to measure the morphological parameters,
while $R_{Abell}$ samples only the brightest galaxy
members. Furthermore, if the core is mainly produced by the fainter
galaxies, as suggested by AMUS01, we expect a better correlation with
$R_{LF}$ than with $R_{Abell}$.  This could also explain the lack of
correlation found by G95, who used the Abell richness.

The better reliability of the richness $R_{LF}$ could also be supported by
the mild correlation of the cluster X-ray luminosity with $R_{LF}$ (correlation coefficient 0.6),
even with only 21 clusters for which X-ray luminosities are available, while no evident correlation 
can be seen with the Abell richness (see figure \ref{fig:richlflx}).

Finally, we checked for correlations between the ellipticity and the
cluster richness, as found for instance by de Theije et al. (1995). In
figure \ref{fig:ellrich} we plot the ellipticity-richness relation for
both $R_{LF}$ and $R_{A}$; empty symbols mark the less regular
clusters discussed earlier, that are expected to show a larger
ellipticity. We do not find any significant correlation between the
ellipticity and the cluster richness. We note however that clusters
with $R_{A}<40$ tend to reach higher ellipticities, while no cluster
with $\epsilon>0.6$ is found at larger richnesses. Thus we cannot
exclude that richer clusters tend to have, on average, smaller
ellipticities. A larger sample is required to confirm this trend.

No evident correlation can be seen between the cluster 
morphological parameters and the cluster X-ray luminosity; however, 
X-ray luminosities are available for only 21 clusters of this sample, and the errors 
on the derived morphological parameters are large, thus this work is not suitable
to study such correlations.

\section{Conclusions}
\label{conclusions}
As already stressed by many authors (see, for instance, G95), the
derivation of cluster's morphological parameters is strongly dependent
on the properties of the galaxy sample. In particular, inhomogeneity
of data (in terms of photometric bands, limiting magnitudes, etc.),
and incomplete coverage of the cluster area, may introduce systematic
errors in the derived sizes, richnesses, ellipticities, etc. In order
to minimize these problems, our sample was derived from a homogeneous
data set extracted from DPOSS data with well controlled photometric
errors, limiting magnitudes, and wide area coverage. The sample was
cleaned of all objects with visually evident signs of
substructure, in an attempt to fit only compact clusters for which
meaningful values of the profile parameters could be derived. The
morphological parameters, namely the core radius, $\beta$ power law
index, ellipticity, position angle, and richness were derived, using
software originally designed for X-ray data analysis, from the
unsmoothed and essentially unbinned galaxy distributions.

The most relevant results of this work may be summarized as
follows:
\begin{itemize}

\item One third of the clusters in this sample have core radii smaller
than 100 kpc, which is close to the limit that our data allow us to
resolve, and possibly consistent with cusped profiles.

\item The remaining clusters span a broad range of core radii up to
$\sim 1500$ kpc. While a few of these clusters are likely to be
disturbed systems, we find several objects with regular morphology
which seem to possess a well resolved core radius.

\item More than 80 per cent of this compact cluster sample has
ellipticity greater than $0.2$, with an average ellipticity of $\sim
0.5$.

\item The comparison of the cluster position angles obtained within
1.5 and 3 Mpc are in very good agreement, thus confirming the absence
of large substructures, and the lack of significant twisting of the
cluster isopleths. We confirm the strong alignment between the
cluster position angle and the major axis of the brightest ellipticals
close to the cluster centre.

\item We find an average power-law slope of the cluster profiles
$\overline{\beta}\sim 0.8$). This result could be compatible with flat
universe models either with zero (standard CDM) or non-zero
cosmological constant. 

\item We find evidence that the core radius is correlated, albeit
with large scatter, with the cluster richness, if the entire cluster
population is taken into account. On the other hand, we do not find
any significant correlation between cluster ellipticity and richness,
although a mild trend cannot be excluded with our data.

\end{itemize}

{Future work will allow to confirm and extend our results, using the
procedure developed in the present study, to estimate morphological
parameters for the larger cluster sample now available in the DPOSS
catalogs. Furthermore, it would be desirable to apply the same
procedure to galaxy clusters extracted from cosmological simulations
and compare the results with real clusters, to derive more accurate
information about the cosmological parameters and the process through
which galaxy clusters are assembled.}

\section*{Acknowledgments}
The authors wish to thank Andrea Biviano, together with Christophe Adami, 
for kindly providing us with his code at the beginning of this study, and for several 
helpful suggestions on its use. We also thank Elisabetta De
Filippis for many useful discussions. VS acknowledges helpful discussions 
with Gabriella De Lucia, Maurilio Pannella and Paola Popesso.
We thank the referee for very helpful and constructive comments.
VS warmly thanks Massimo Capaccioli for assistance and hospitality
at the Osservatorio Astronomico di Capodimonte, and is grateful to Ralf Bender and all his group
for hospitality at MPE during part of this work.
This work was partly funded by the Italian Ministry of Research through a COFIN grant and by the
European Social Fund (FSE) through a Ph.D. grant. 
Processing and cataloging of DPOSS
was supported by a generous grant from the Norris Foundation,
and by other private donors.
This research has made use of the X-Rays Clusters Database (BAX)
which is operated by the Laboratoire d'Astrophysique de Tarbes-Toulouse (LATT),
under contract with the Centre National d'Etudes Spatiales (CNES).

\begin{table*}
\centering
\begin{minipage}{140mm}
\caption{The compact cluster sample. The second and third columns give the cluster coordinates (from the Abell catalogue) at equinox J2000.0. In columns 4, 5, 6, and 7 we give, respectively, the cluster spectroscopic redshift derived from literature (in parentheses, the number N$_{z}$ of galaxies used to measure $z$; see \citet{strublerood99} for discussion on lower limits), the Abell richness class (Abell 1958), the X-ray luminosity in the 0.1-2.4 keV band (from the BAX database - LATT), and the plate of the POSS archive where the cluster is imaged.   \label{tab:compactsample}}
    \begin{tabular}{@{}l l l l l l l@{}}
\hline
 Abell N.&RA&Dec&z (N$_{z}$) &Abell Richness Class&$L_{X}$ ($10^{44}$ erg/s) &Plate\vspace{0.06cm}\\
\hline
28 & 0:25:10 & 8:08:34 & 0.184 ( - ) & 2 &-& 680\vspace{0.06cm}\\
41 & 0:28:46 & 7:51:36 & 0.275 ( 3 ) & 3 &-& 752\vspace{0.06cm}\\
79 & 0:40:37 & 18:08:27 & 0.093 ( 1 ) & 1 &-& 540\vspace{0.06cm}\\
84 & 0:41:50 & 21:24:25 & 0.103 ($>$1 ) & 1 &1.82& 540\vspace{0.06cm}\\
98 & 0:46:26 & 20:29:24 & 0.104 (24 ) & 3 &0.94& 540\vspace{0.06cm}\\
171 & 1:16:46 & 16:15:46 & 0.070 ($>$4 ) & 0 &-& 611\vspace{0.06cm}\\
175 & 1:19:33 & 14:52:44 & 0.129 ($>$0 ) & 2 &-& 611\vspace{0.06cm}\\
192 & 1:24:17 & 4:29:38 & 0.121 ( - ) & 2 &0.84& 755\vspace{0.06cm}\\
286 & 1:58:26 & -1:46:26 & 0.160( - ) & 2 &2.53& 829\vspace{0.06cm}\\
566 & 7:04:29 & 63:17:31 & 0.098 ( 9 ) & 2 &2.81& 088\vspace{0.06cm}\\
655 & 8:25:20 & 47:08:13 & 0.124 ( 1 ) & 3 &4.90& 210\vspace{0.06cm}\\
763 & 9:12:28 & 16:00:36 & 0.085 ( 5 ) & 1 &2.27& 634\vspace{0.06cm}\\
910 & 10:02:59 & 67:10:30 & 0.206 ( 2 ) & 4 &5.43& 091\vspace{0.06cm}\\
971 & 10:19:46 & 40:58:55 & 0.06  ( ? ) & 1 &1.10& 317\vspace{0.06cm}\\
1661 & 13:01:48 & 29:04:51 & 0.167 ( 4 ) & 2 &0.65& 443\vspace{0.06cm}\\
1672 & 13:04:45 & 33:33:57 & 0.188 ( 1 ) & 1 &4.25& 382\vspace{0.06cm}\\
1677 & 13:05:52 & 30:53:56 & 0.184 ( 1 ) & 2 &5.37& 443\vspace{0.06cm}\\
1835 & 14:01:01 & 2:51:32 & 0.252 ($>$1 ) & 0 &29.80& 793\vspace{0.06cm}\\
1902 & 14:21:46 & 37:18:21 & 0.16  ( 1 ) & 2 &4.55& 326\vspace{0.06cm}\\
1914 & 14:26:02 & 37:49:33 & 0.171 ( 2 ) & 2 &17.30& 326\vspace{0.06cm}\\
2061 & 15:21:15 & 30:39:18 & 0.077 (20 ) & 1 &4.85& 449\vspace{0.06cm}\\
2065 & 15:22:42 & 27:43:22 & 0.072 (22 ) & 2 &5.55& 449\vspace{0.06cm}\\
2069 & 15:23:57 & 29:54:25 & 0.116 ( 9 ) & 2 &3.45& 449\vspace{0.06cm}\\
2083 & 15:29:26 & 30:44:45 & 0.114 ( 1 ) & 1 &-& 449\vspace{0.06cm}\\
2142 & 15:58:16 & 27:13:30 & 0.091 (103) & 2 &21.24& 516\vspace{0.06cm}\\
2177 & 16:20:58 & 25:44:56 & 0.161 ($>$0 ) & 0 &2.20& 517\vspace{0.06cm}\\
2178 & 16:21:30 & 24:39:00 & 0.093 ( 2 ) & 1 &0.28& 517\vspace{0.06cm}\\
2223 & 16:42:30 & 27:26:24 & 0.103 ( - ) & 0 &-& 517\vspace{0.06cm}\\
\hline
     \end{tabular}
\end{minipage}

\end{table*}

\begin{table*}
\centering
 \begin{minipage}{140mm}
    \caption{Parameter estimates for the compact clusters. Column 1: cluster identification; columns 2-6: 
    estimated parameters: core radius, $\beta$ power law index, ellipticity, major axis position angle and projected central density. The errors are 1$\sigma$ levels derived as discussed in the text. The tabulated 
$\Sigma_{0}$ are not corrected for the different absolute completeness limits of the clusters. \label{tab:results1}}
   \begin{tabular}{@{}l l l l l l@{}}
    \hline
  Abell N.&    $ r_{core}$ &	$ \beta$&$ \epsilon$&$ \theta$&$\Sigma_{0}^{*}$\vspace{0.06cm}\\
 & (kpc)& & &(rad)&$10^{-5}$gal./kpc$^{2}$\vspace{0.06cm}\\
\hline
  28  	&$  370  _{- 150}^{+ 260}$	&$   1.2  _{-0.3}^{+  0.5}$    &$  0.61 _{-0.07 }^{+0.06 }$      &$  0.06  _{-0.15}^{+0.17}$	    &$ 9 _{-3}^{+6}  $\vspace{0.06cm}\\
  41  	&$ 1000  _{- 370}^{+2200}$	&$   1.6  _{-0.5}^{+  6}$    &$  0.71 _{-0.05 }^{+0.05 }$        &$  1.27  _{-0.07}^{+0.07}$	    &$ 4.6 _{-1.4}^{+ 1.8}  $\vspace{0.06cm}\\
  79  	&$  140  _{-  89}^{+  96}$	&$  0.83  _{-0.15}^{+  0.2}$    &$    0.55 _{-0.1 }^{+0.09 }$    &$   1.78  _{-0.2}^{+0.16}$	    &$ 25 _{-9}^{+ 40}  $\vspace{0.06cm}\\
  84  	&$  360  _{- 140}^{+ 210}$	&$  0.96  _{-0.16}^{+  0.2}$    &$  0.47 _{-0.09 }^{+0.08 }$     &$  2.21  _{-0.19}^{+0.2}$	    &$ 12 _{-4}^{+ 7}  $\vspace{0.06cm}\\
  98  	&$  970  _{- 300}^{+ 490}$	&$  0.97  _{-0.18}^{+  0.3}$    &$  0.52 _{-0.04 }^{+0.04 }$     &$   2.94  _{-0.09}^{+0.09}$	    &$ 8.1 _{-1.7}^{+ 2}  $\vspace{0.06cm}\\
 171  	&$  160  _{-  63 }^{+ 94 }$    &$  1.00  _{-0.16}^{+  0.3}$    &$  0.25 _{-0.14}^{+0.13}$        &$  1.9  _{-0.5}^{+0.8}$	    &$ 27 _{-9}^{+15}  $\vspace{0.06cm}\\
 175  	&$  390  _{- 160 }^{+ 210}$     &$  0.70  _{-0.09}^{+ 0.13}$    &$  0.58 _{-0.05}^{+0.05 }$      &$  2.34  _{-0.11}^{+0.12}$	    &$ 9 _{-2}^{+5}  $\vspace{0.06cm}\\
 192  	&$   33  _{-  30}^{+  57}$	&$  0.79  _{-0.1}^{+ 0.15}$    &$  0.73 _{-0.07}^{+0.07 }$       &$  2.71  _{-0.17}^{+0.14}$	    &$ 134 _{-80}^{+\infty}  $\vspace{0.06cm}\\
 286  	&$  310  _{- 120}^{+ 210}$	&$  0.94  _{-0.14}^{+  0.3}$    &$  0.36 _{-0.09 }^{+0.09 }$     &$  3.1  _{-0.2}^{+0.2}$	    &$ 11 _{-4}^{+ 6}  $\vspace{0.06cm}\\
 566  	&$   100  _{-  67 }^{+69}$	&$  0.44  _{-0.04}^{+ 0.04}$     &$  0.17 _{-0.09 }^{+0.08}$     &$  2.3  _{-0.6}^{+0.5}$	    &$ 19 _{-6}^{+ 20}  $\vspace{0.06cm}\\
 655  	&$ 1300  _{- 410}^{+ 550}$	&$   1.3  _{-0.3}^{+  0.5}$    &$  0.34 _{-0.04 }^{+0.04 }$      &$  1.62  _{-0.12}^{+0.12}$	    &$ 6.8 _{-1}^{+ 1.5}  $\vspace{0.06cm}\\
 763  	&$   30  _{-  29}^{+  35}$	&$  0.61  _{-0.07}^{+  0.07}$    &$  0.54 _{-0.09 }^{+0.09  }$   &$ 0.9  _{-0.2}^{+0.2}$	    &$ 84 _{-50}^{+ \infty}  $\vspace{0.06cm}\\
 910  	&$ 1400  _{- 520}^{+2600}$	&$  1.0  _{-0.2}^{+  1.8}$    &$  0.44 _{-0.07  }^{+0.06 }$      &$  2.97  _{-0.18}^{+0.17}$	    &$ 1.9 _{-0.6}^{+ 0.5}  $\vspace{0.06cm}\\
 971  	&$   39  _{-  37}^{+  34}$	&$  0.46  _{-0.03}^{+ 0.04}$    &$  0.70 _{-0.03 }^{+0.03 }$     &$  1.50 _{-0.07}^{+0.07}$	    &$ 91 _{-40}^{+\infty}  $\vspace{0.06cm}\\
1661	&$  890  _{- 400 }^{+ 770 }$    &$   1.1  _{-0.3 }^{+  0.8}$    &$  0.64 _{-0.06 }^{+0.05 }$	 &$  1.29  _{-0.10 }^{+0.10 }$      &$ 4.2 _{-1.3}^{+2}  $\vspace{0.06cm}\\
1672  	&$  280  _{- 150 }^{+ 270 }$    &$  0.74  _{-0.15}^{+  0.3}$    &$0.006 _{-0.006}^{+0.19  }$	 &$  2.4  _{-2.4}^{+4}$		    &$ 3.4 _{-1.3}^{+3}  $\vspace{0.06cm}\\
1677  	&$   15  _{-  14 }^{+  24 }$    &$   0.57  _{-0.07}^{+ 0.07}$    &$  0.43 _{-0.14}^{+0.12  }$	 &$  2.6  _{-0.4}^{+0.3}$	    &$ 73 _{-50}^{+\infty}  $\vspace{0.06cm}\\
1835  	&$   81  _{-  58}^{+ 130}$	&$  0.54  _{-0.07}^{+ 0.07}$    &$  0.56 _{-0.07}^{+0.06 }$      &$  2.58  _{-0.16}^{+0.15}$	    &$ 19 _{-10}^{+\infty}  $\vspace{0.06cm}\\
1902  	&$  380  _{- 170}^{+ 250}$	&$  0.66  _{-0.11}^{+  0.16}$	&$  0.1  _{-  0.1 }^{+ 0.16}$    &$    2.5  _{-2 }^{+0.6}$	    &$ 4.6 _{-1.3}^{+2}  $\vspace{0.06cm}\\
1914  	&$  140  _{-  80}^{+  88}$	&$  0.63  _{-0.08}^{+ 0.07}$    &$  0.50 _{-0.07}^{+0.06 }$      &$ 0.99  _{-0.15}^{+0.15}$	    &$ 18 _{-7}^{+20}  $\vspace{0.06cm}\\
2061  	&$  780  _{- 300}^{+ 260}$	&$  0.87  _{-0.16}^{+  0.16}$    &$  0.70 _{-0.03}^{+0.03 }$	 &$ 0.78  _{-0.06}^{+0.06}$	    &$ 10.7 _{-1.9}^{+4}  $\vspace{0.06cm}\\
2065  	&$   97  _{-  51}^{+  48}$	&$  0.60  _{-0.05}^{+ 0.05}$    &$  0.17 _{-0.08 }^{+0.08 }$     &$  2.8  _{-0.5}^{+0.3}$	    &$ 48 _{-14}^{+40}  $\vspace{0.06cm}\\
2069  	&$  340  _{- 150}^{+ 190}$	&$  0.82  _{-0.15}^{+  0.19}$    &$  0.35 _{-0.09 }^{+0.08 }$    &$  2.7  _{-0.3}^{+0.3 }$	    &$ 9 _{-3}^{+5}  $\vspace{0.06cm}\\
2083  	&$   23  _{-  18}^{+  20}$	&$  0.64  _{-0.06}^{+ 0.05}$    &$  0.47 _{-0.08 }^{+0.08 }$     &$  1.4  _{-0.2}^{+0.2}$	    &$ 140 _{-70}^{+\infty}  $\vspace{0.06cm}\\
2142  	&$  857  _{- 310}^{+ 300}$	&$  0.79  _{-0.14}^{+  0.15}$    &$  0.54 _{-0.04 }^{+0.03 }$    &$  2.46  _{-0.09}^{+0.1}$	    &$ 8.5 _{-1.4}^{+3}  $\vspace{0.06cm}\\
2177  	&$   55  _{-  50}^{+ 130}$	&$  0.53  _{-0.08}^{+ 0.08}$    &$  0.72 _{-0.07 }^{+0.06 }$     &$  1.93  _{-0.13}^{+0.15}$	    &$ 25 _{-16}^{+\infty}  $\vspace{0.06cm}\\
2178  	&$  340  _{- 130}^{+ 440}$	&$  0.86  _{-0.14}^{+  0.5}$    &$    0.40 _{-0.11 }^{+0.1  }$   &$  1.4  _{-0.2}^{+0.2}$	    &$ 11 _{-4}^{+5}  $\vspace{0.06cm}\\
2223  	&$   37  _{-  29}^{+  53}$	&$  0.60  _{-0.06}^{+ 0.06}$    &$  0.60 _{-0.07 }^{+0.06 }$     &$  1.15  _{-0.13}^{+0.13}$	    &$ 78 _{-40}^{+\infty} $\vspace{0.06cm}\\
  \hline
     \end{tabular}

\end{minipage}  
\end{table*}

\begin{table*}
\centering
 \begin{minipage}{140mm}
    \caption{Parameter estimates for the compact clusters. Column 1: cluster identification; columns 2-3: richness estimates ($R_{A}$, $R_{LF}$) obtained as described in the text; column 4: total number of galaxies used for profile fitting in the $6 \times 6$ Mpc$^{2}$ region; column 5: background density as estimated in the external annulus ($4 \leq r \leq 5$ Mpc); column 6: background density as estimated with a simultaneous fit on all 8 parameters; column 7: Kolmogorov Smirnov 
confidence levels for the distribution of the cluster galaxy position angles. The errors are 1$\sigma$ levels. \label{tab:results2}}
   \begin{tabular}{@{}l l l l l l l@{}}
    \hline
  Abell N.&  $R_{A} $&$R_{LF} $&$N(6 \times 6$ Mpc$^{2})$ & $\Sigma_{bkg}$&$\Sigma_{bkg}^{fit}$&$P_{KS}(d>d_{obs})$\vspace{0.06cm}\\
 & (gal.)&(gal.)&(gal.)& gal/sqarcmin &gal/sqarcmin&\vspace{0.06cm}\\
\hline
  28  	&  $   39	$&$  69$&	     	306&$ 0.36 \pm 0.03$&0.33& $0.8$\vspace{0.06cm}\\
  41  	&  $   35 	$&$ 216$&		189&$ 0.33 \pm 0.04$&0.35& $  0.3$\vspace{0.06cm}\\
  79  	&  $   28 	$&$38$&          		632&$ 0.34 \pm 0.02$&0.30& $0.4$\vspace{0.06cm}\\
  84  	&  $   36 	$&$  81$&	     	678&$ 0.42 \pm 0.02$&0.37& $0.5$\vspace{0.06cm}\\
  98  	&  $   50	$&$ 326$&	  	853&$ 0.31 \pm 0.02$&0.30& $0.4$\vspace{0.06cm}\\
 171  	&  $   23 	$&$  46$&	     	939&$ 0.30 \pm 0.01$&0.30& $ 0.2$\vspace{0.06cm}\\
 175  	&  $   39 	$&$  86$&	     	532&$ 0.28 \pm 0.02$&0.28& $ 0.5$\vspace{0.06cm}\\
 192  	&  $   33 	$&$  80$&	       	337&$ 0.28 \pm 0.02$&0.24& $ 0.7$\vspace{0.06cm}\\
 286  	&  $   26 	$&$  140$&	    	327&$ 0.29 \pm 0.02$&0.31& $ 0.7$\vspace{0.06cm}\\
 566  	&  $   57 	$&$ 108$&	    	834&$ 0.36 \pm 0.02$&0.38& $0.3$\vspace{0.06cm}\\
 655  	&  $   65 	$&$ 329$&	  	630&$ 0.31 \pm 0.03$&0.28& $0.4$\vspace{0.06cm}\\
 763  	&  $   10 	$&$  79$&	       	818&$ 0.33 \pm 0.01$&0.32& $0.04$\vspace{0.06cm}\\
 910  	&  $   49 	$&$  164$&	       	232&$ 0.25 \pm 0.03$&0.17& $0.2$\vspace{0.06cm}\\
 971  	&  $   38 	$&$  74$&	       	1641&$ 0.35 \pm 0.01$&0.34& $ 0.05$\vspace{0.06cm}\\
1661	&  $   39 	$&$ 104$&	   	326&$ 0.30 \pm 0.02$&0.31& $0.02$\vspace{0.06cm}\\
1672  	&  $   30 	$&$  76$&	   	217&$ 0.26 \pm 0.02$&0.23& $  0.5$\vspace{0.06cm}\\
1677  	&  $   30 	$&$125$&	           	245&$ 0.22 \pm 0.02$&0.22& $  0.1$\vspace{0.06cm}\\
1835  	&  $   64 	$&$  214$&	       	274&$ 0.25 \pm 0.03$&0.23& $ 0.02$\vspace{0.06cm}\\
1902  	&  $   39 	$&$  183$&	  	366&$ 0.26 \pm 0.02$&0.26& $0.3$\vspace{0.06cm}\\
1914  	&  $   43 	$&$  171$&	    	370&$ 0.31 \pm 0.02$&0.31& $0.3$\vspace{0.06cm}\\
2061  	&  $   37 	$&$231$&		056&$ 0.33 \pm 0.01$&0.33& $0.4$\vspace{0.06cm}\\
2065  	&  $   67 	$&$228$&		583&$ 0.34 \pm 0.01$&0.35& $0.005$\vspace{0.06cm}\\
2069  	&  $   22 	$&$  162$&	     	625&$ 0.38 \pm 0.02$&0.37& $0.02$\vspace{0.06cm}\\
2083  	&  $   21 	$&$ 93$&        		513&$ 0.29 \pm 0.02$&0.28& $0.2$\vspace{0.06cm}\\
2142  	&  $   59 	$&$  194$&	  	841&$ 0.34 \pm 0.02$&0.28& $ 9 \times 10^{-6}$\vspace{0.06cm}\\
2177  	&  $   27 	$&$95$&	          	468&$ 0.53 \pm 0.03$&0.48& $ 0.3$\vspace{0.06cm}\\
2178  	&  $   19 	$&$  83$&	     	908&$ 0.47 \pm 0.02$&0.45& $0.002$\vspace{0.06cm}\\
2223  	&  $   21 	$&$  73$&		635&$ 0.36 \pm 0.03$&0.36& $0.004$\vspace{0.06cm}\\
  \hline
     \end{tabular}

\end{minipage}  
\end{table*}

\clearpage

\clearpage

%\clearpage

\hspace{10cm}

\begin{figure*}    
\includegraphics[clip,width=17cm,bb= 49 71 530 732]{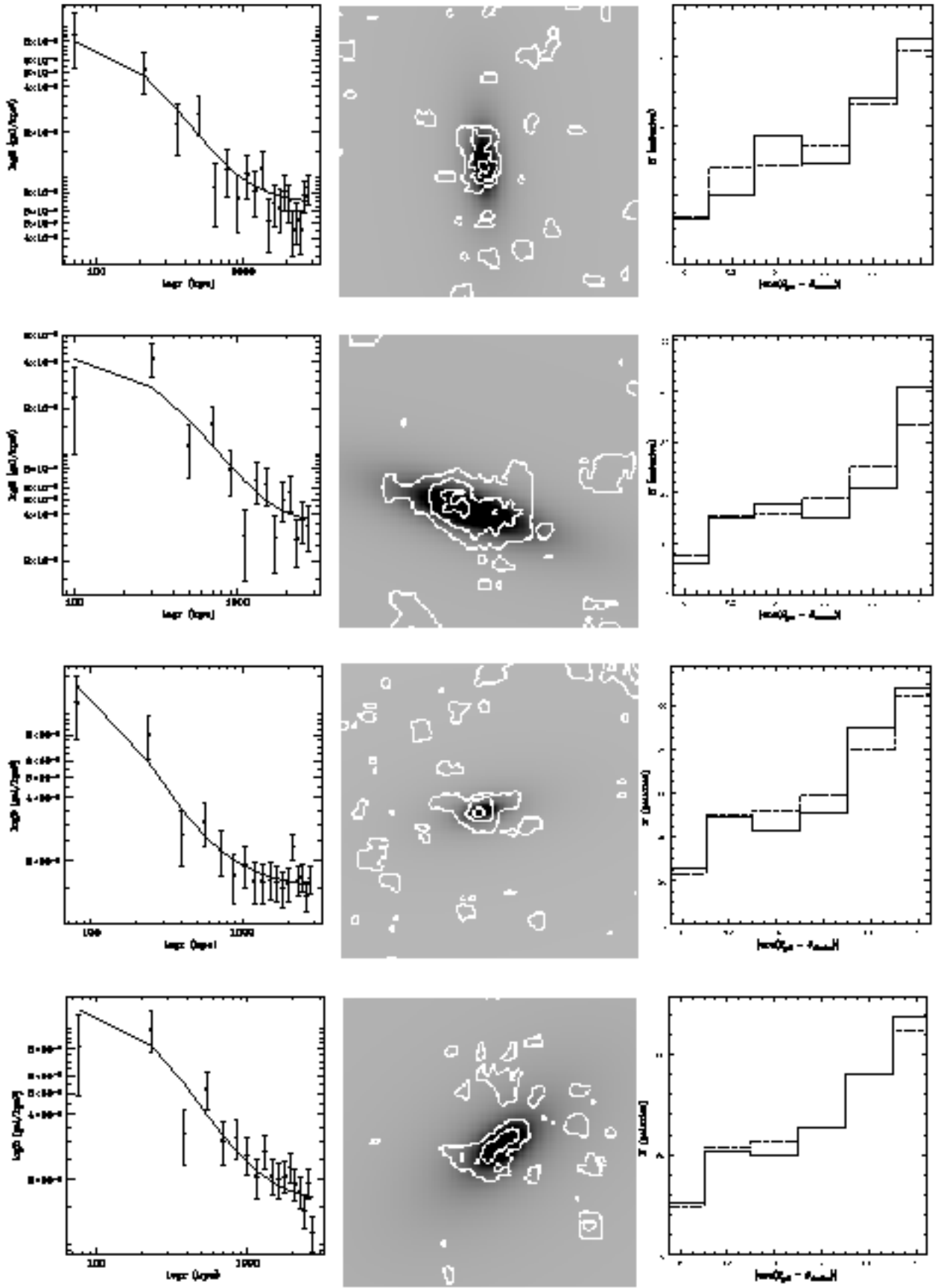}
\caption{Radial profile and contour plot in the fitted region ($6
\times 6$ Mpc$^{2}$ - model images are all shown with the same
contrast/bias) and galaxy P.A. distribution for (starting from top):
A28, A41, A79, A84 (see fig. 2 for details). \label{fig:indivcluster1} }
\end{figure*}  
\addtocounter{figure}{-1}
\begin{figure*}    
\includegraphics[clip,width=17cm,bb= 49 71 530 732]{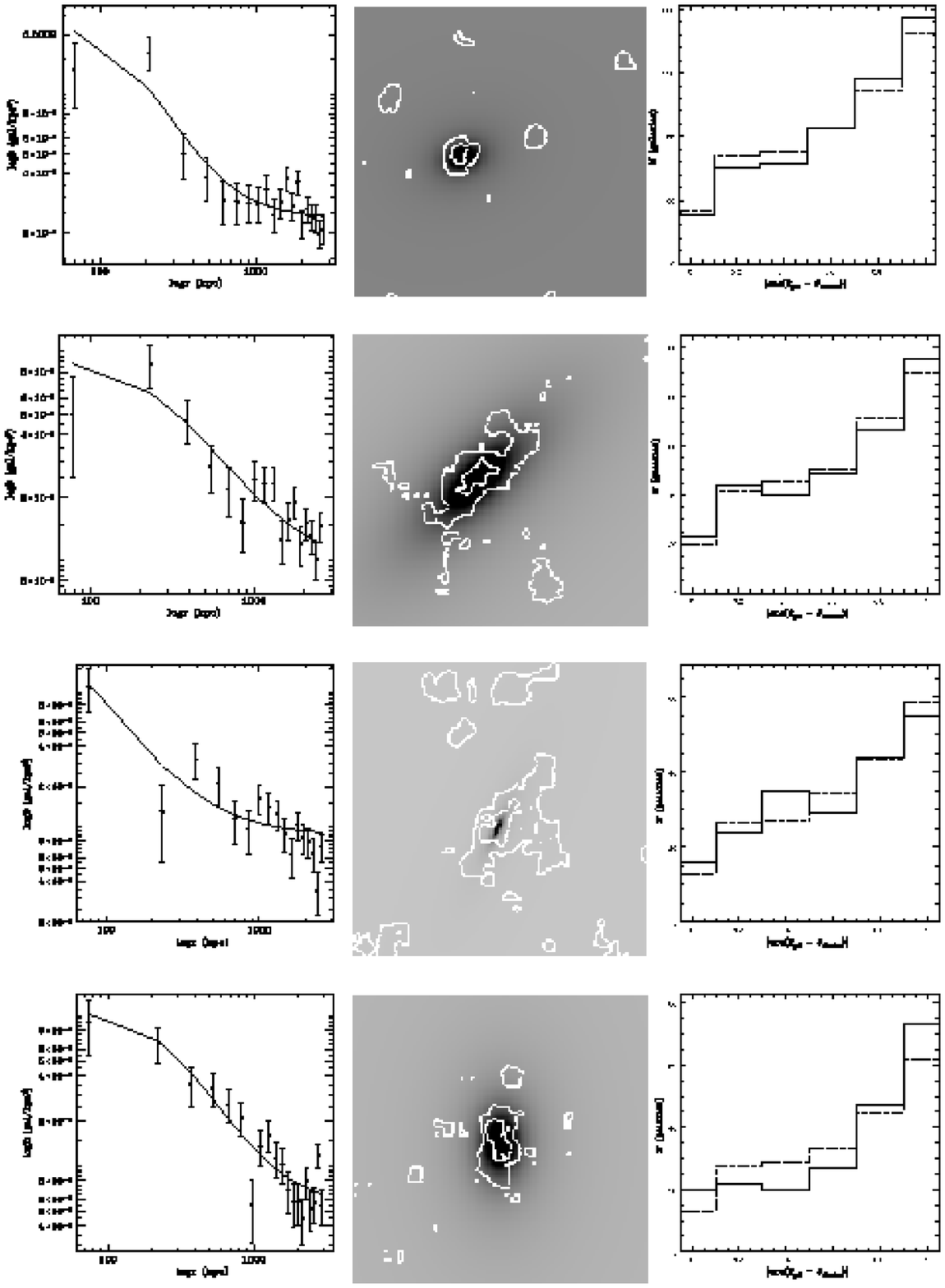}
\caption{(continued). Starting from top: A171, A175, A192, A286.  }  
\end{figure*}  
\addtocounter{figure}{-1}
\begin{figure*}    
\includegraphics[clip,width=17cm,bb= 49 71 530 732]{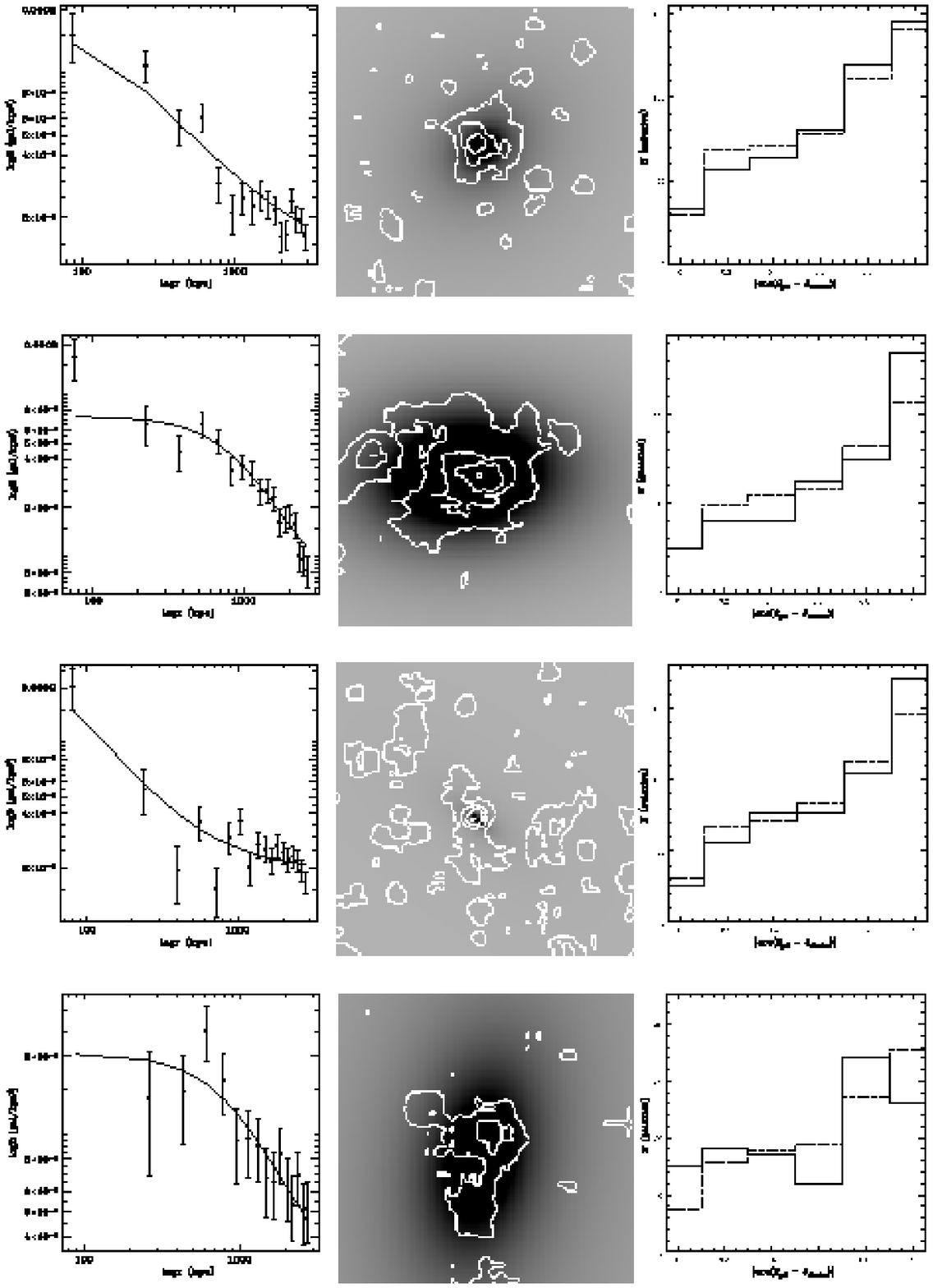}
\caption{(continued). Starting from top: A566, A655, A763, A910.  }   
\end{figure*}  
\addtocounter{figure}{-1}
\begin{figure*}    
\includegraphics[clip,width=17cm,bb= 49 71 530 732]{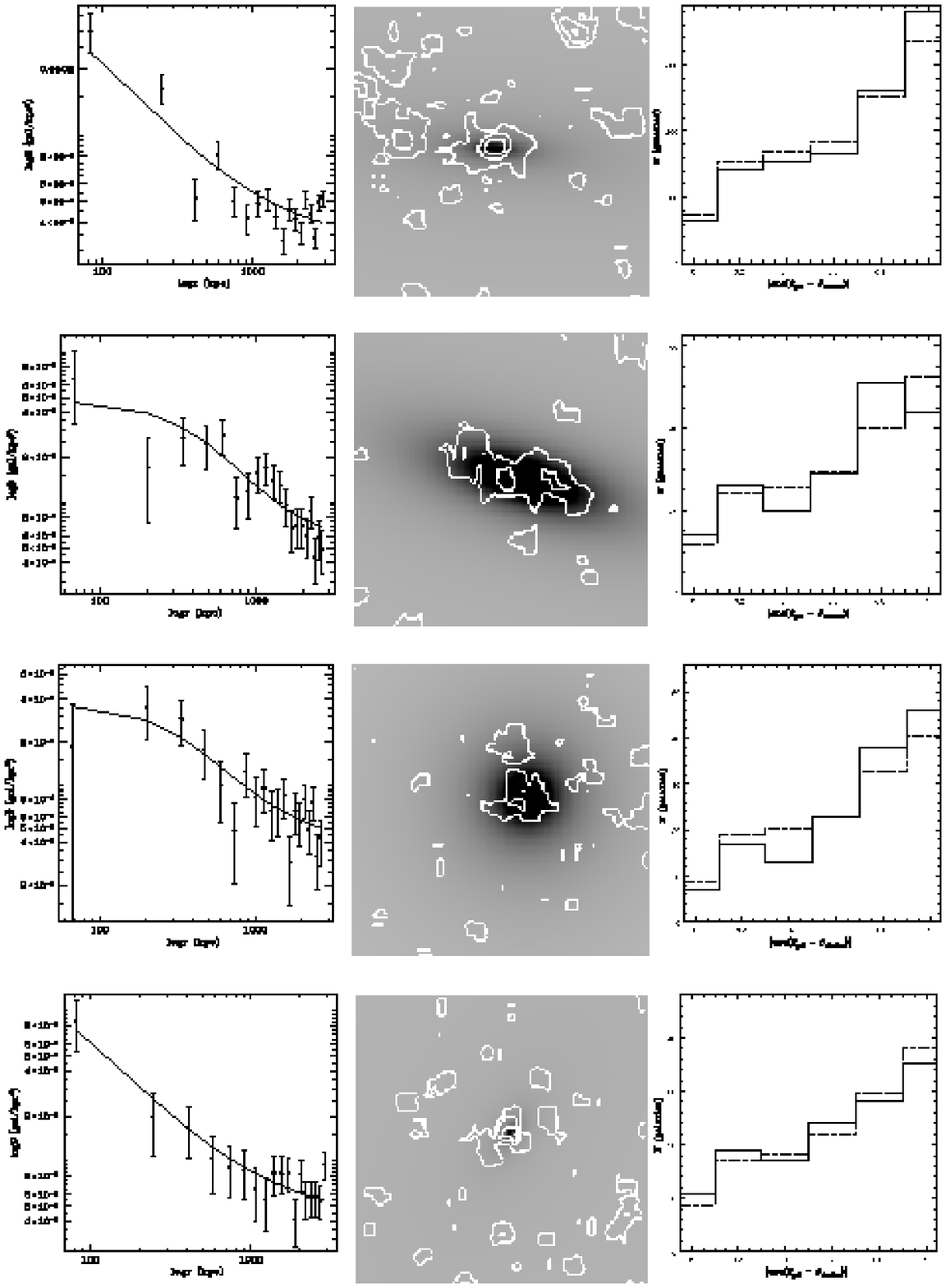}
\caption{(continued). Starting from top: A971, A1661, A1672, A1677.  }   
\end{figure*}  
\addtocounter{figure}{-1}
\begin{figure*}    
\includegraphics[clip,width=17cm,bb= 49 71 530 732]{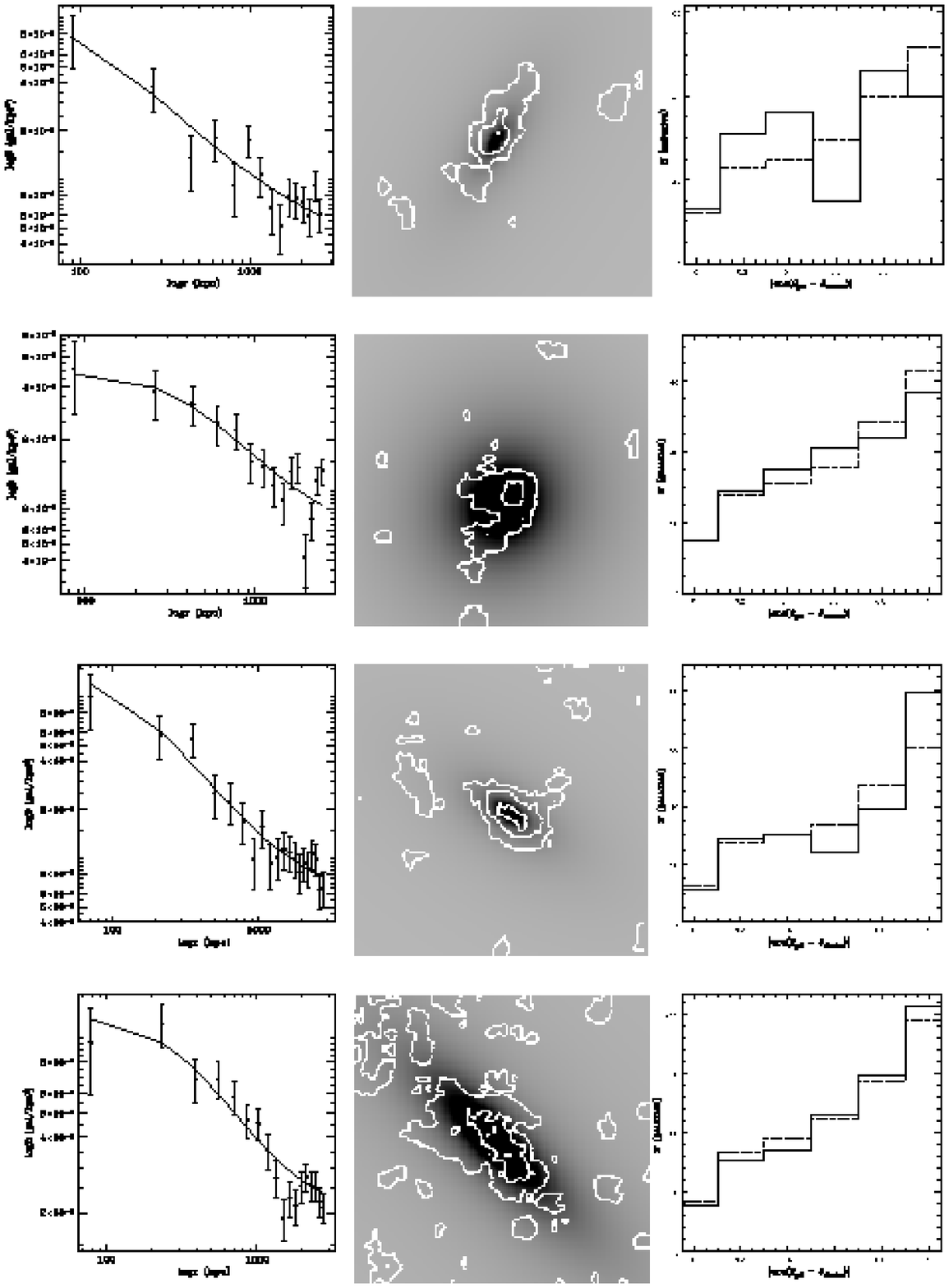}
\caption{(continued). Starting from top: A1835, A1902, A1914, A2061.  }   
\end{figure*}  
\addtocounter{figure}{-1}
\begin{figure*}    
\includegraphics[clip,width=17cm,bb= 49 71 530 732]{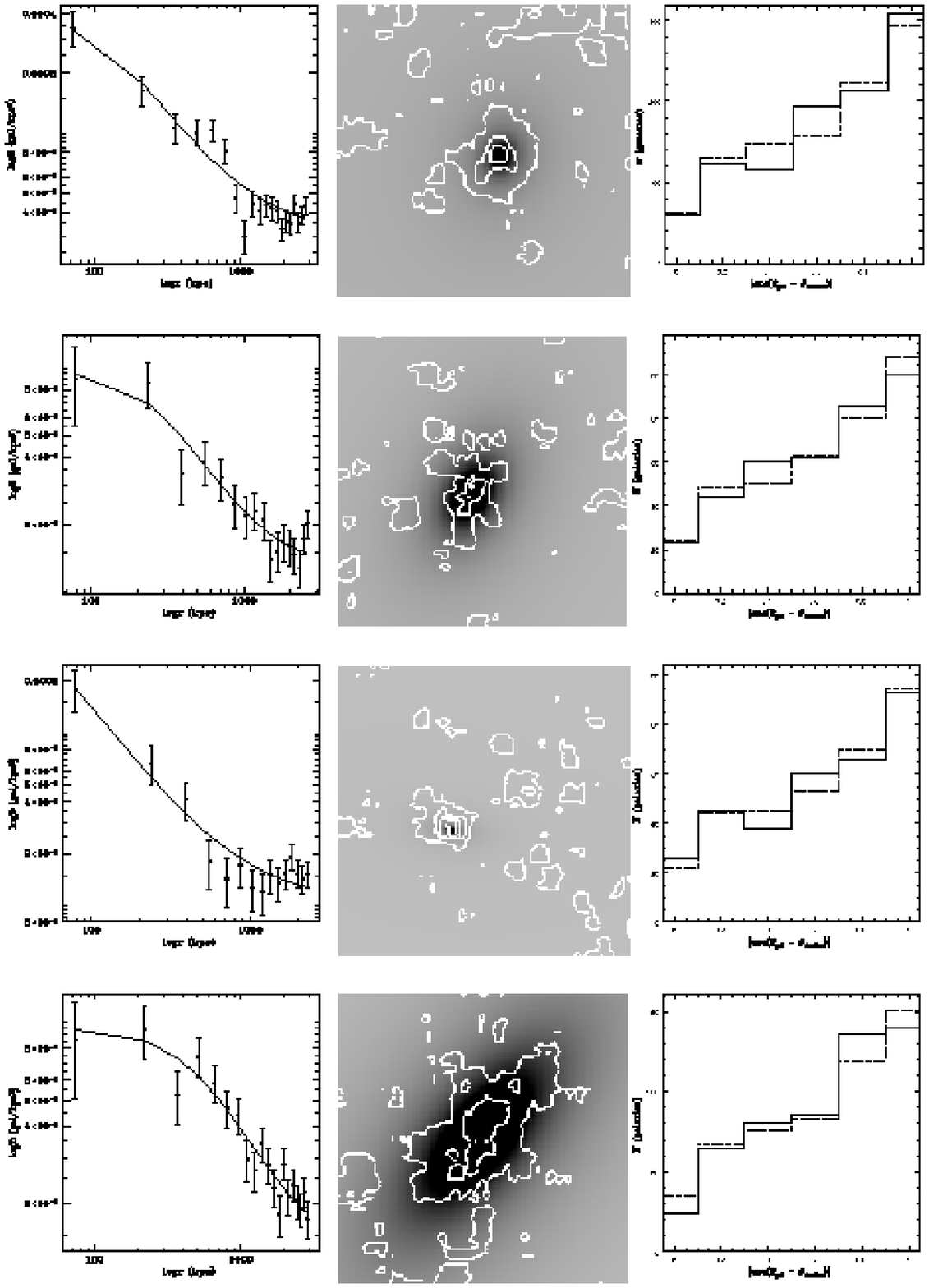}
\caption{(continued). Starting from top: A2065, A2069, A2083, A2142.  }   
\end{figure*}  
\addtocounter{figure}{-1}
\begin{figure*}    
\includegraphics[clip,width=17cm,bb= 49 71 530 732]{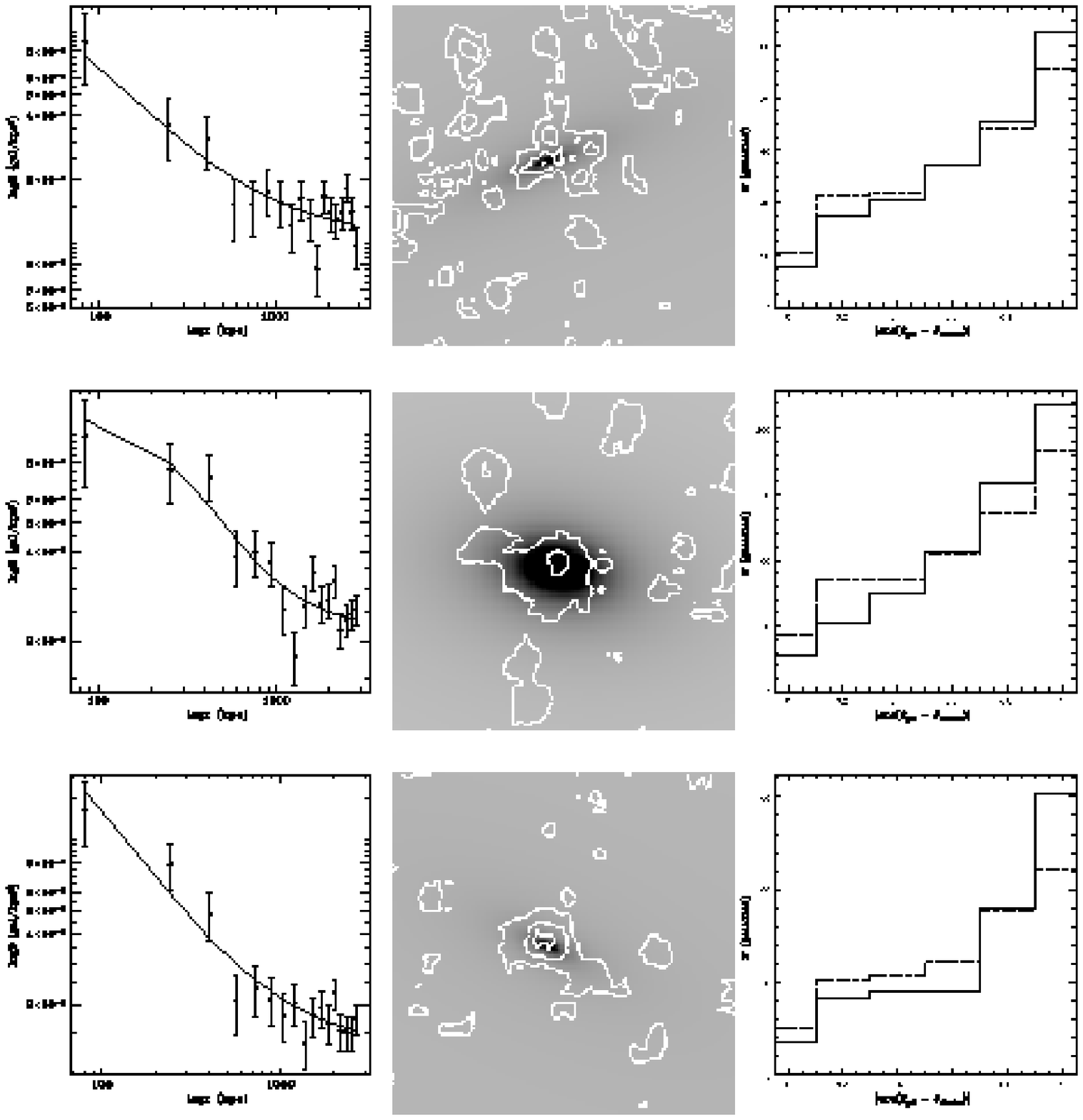}
\caption{(continued). Starting from top: A2177, A2178, A2223.  }   
\end{figure*}

\clearpage

\begin{figure} 
\centerline{\epsfig{file=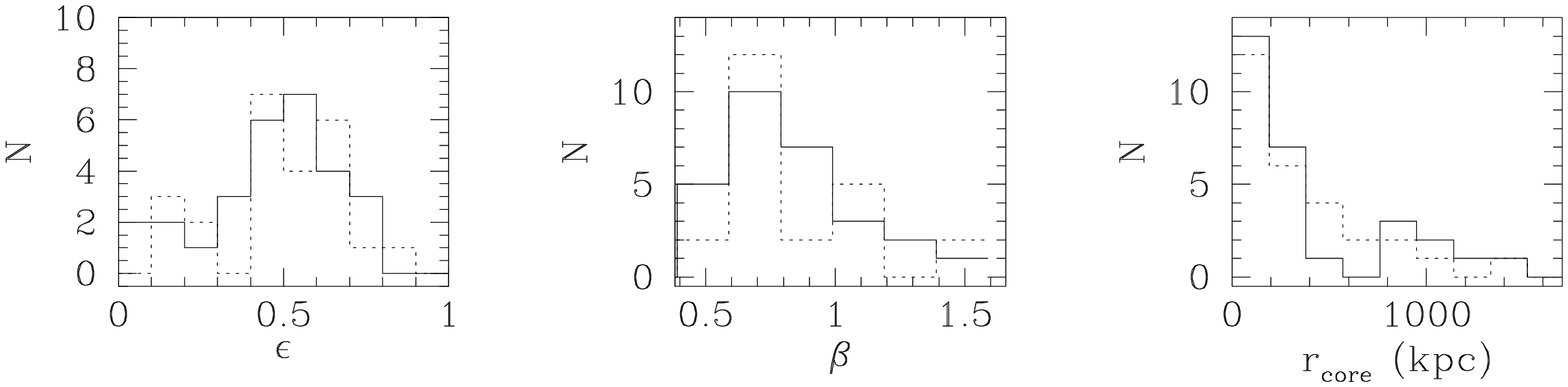,angle=270,width=10cm}}
 \caption{Distributions of the $\beta$-model parameters for the
compact cluster sample: ellipticity (left), $\beta$ power-law index
(center) and core radius (right). The solid and dashed lines show the
results obtained through the fit in the $6\times 6$ Mpc$^{2}$ and
$3\times 3$ Mpc$^{2}$ regions, respectively. \label{fig:histopar}}

 \end{figure}

\begin{figure} 
\epsfxsize=6cm
\centerline{\epsfbox{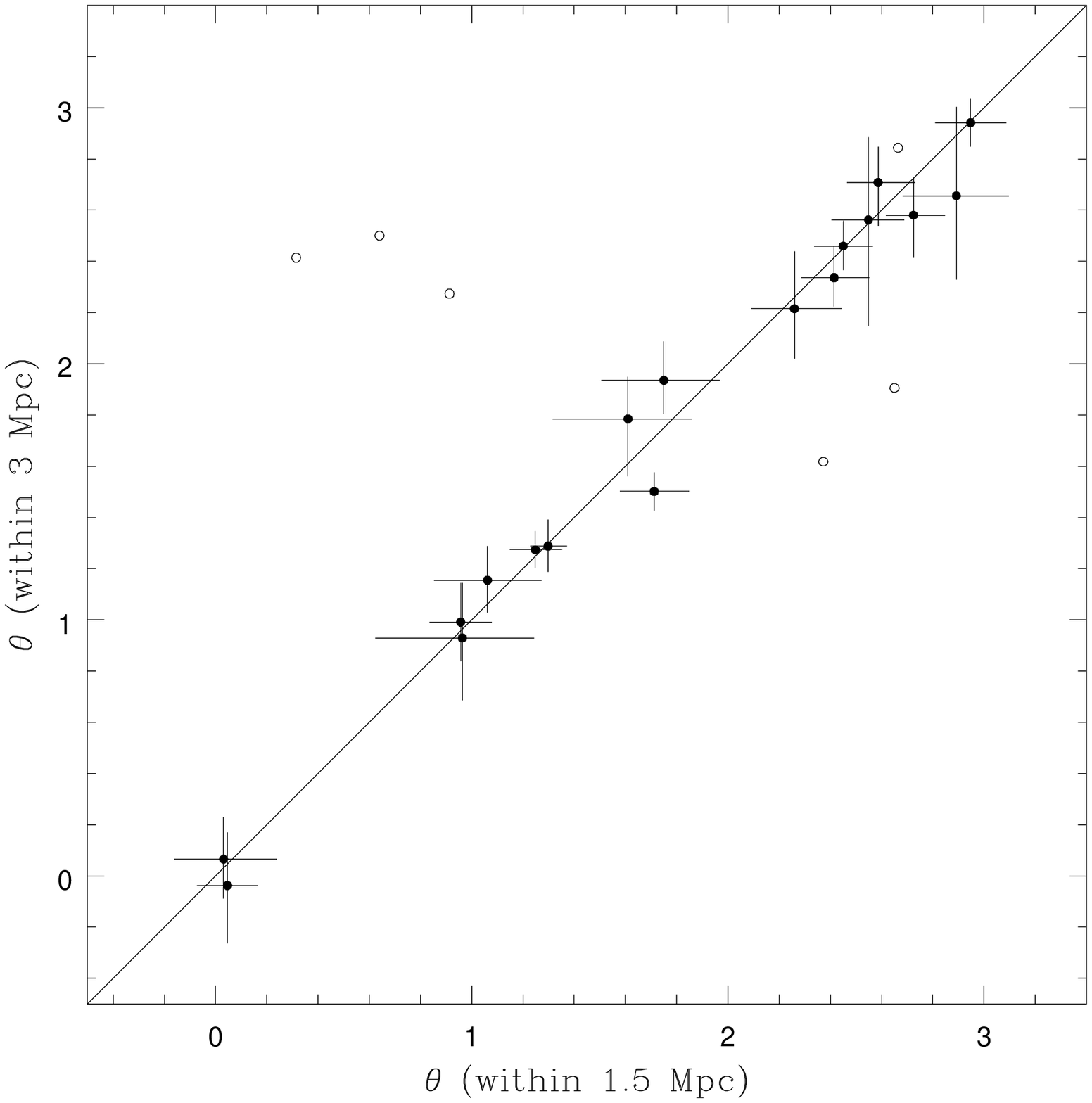}}
 \caption{Position angles obtained for compact clusters fitting the
galaxy distribution within 1.5 Mpc from the cluster centre vs. position
angles within 3 Mpc from cluster centre. The open circles refer to
clusters for which the position angle of the major axis is poorly
determined due to the small ellipticity ($\epsilon < 0.25$); the
continuous line traces the bisector. \label{fig:pa1pa3}}
 \end{figure}

\begin{figure} 
\epsfxsize=6cm
\centerline{\epsfbox{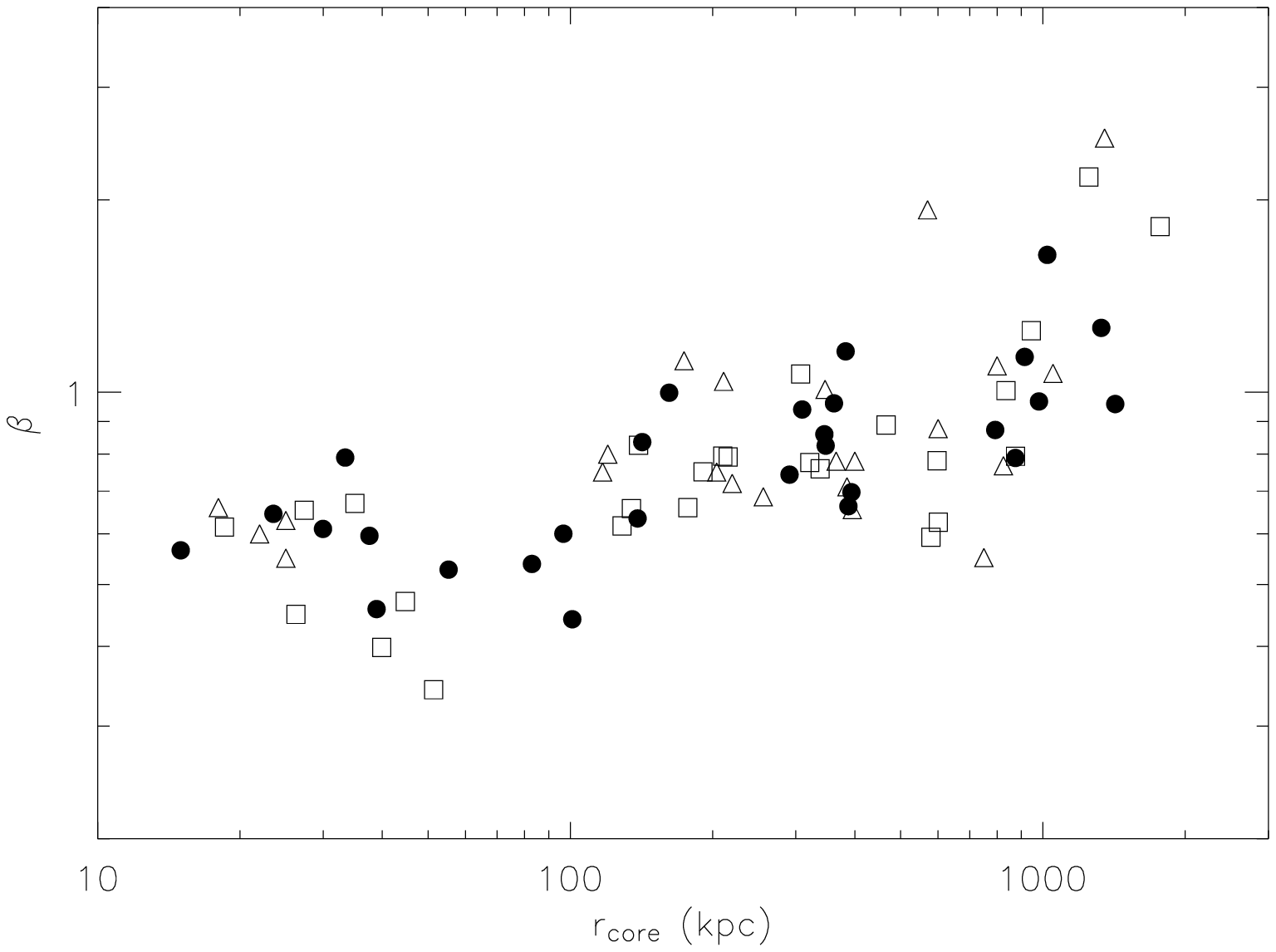}}
 \caption{The coupling of core radius and $\beta$ values. Different symbols show fit results in the three different regions. Solid dots mark the results from the $6 \times 6$ Mpc$^{2}$ region, while triangles and squares mark the results from the $3 \times 3$ and $10 \times 10$ Mpc$^{2}$ regions respectively. \label{fig:corevsbeta}}
 \end{figure}

\begin{figure} 
\epsfxsize=6cm
\centerline{\epsfbox{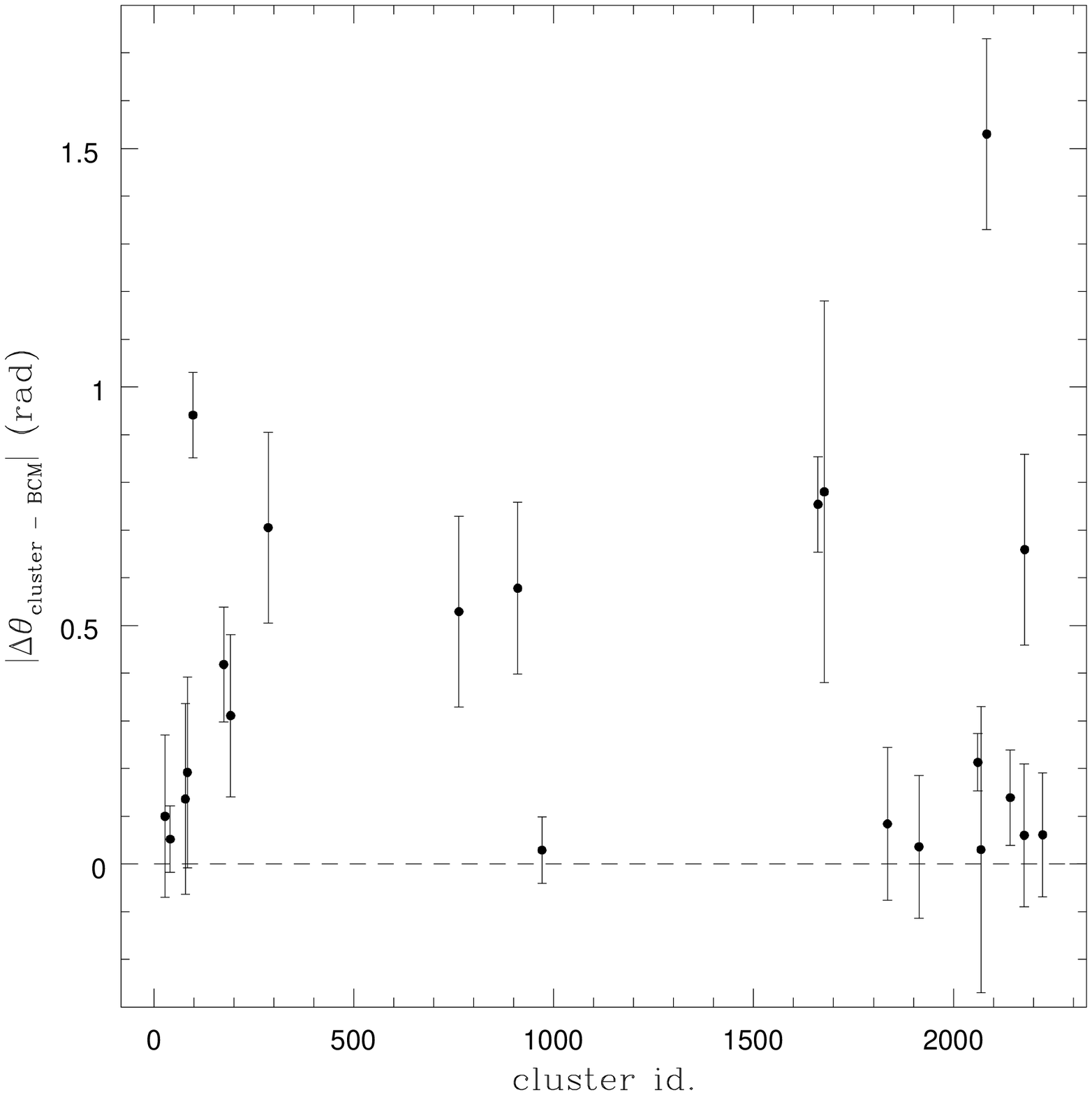}}
 \caption{Orientation of the cluster dominant galaxies relative to the
 cluster major axis (clusters for which the position angle is poorly
 determined (see text and figure \ref{fig:pa1pa3}) are not included). \label{fig:bcm}}

 \end{figure}

\begin{figure} 
\epsfxsize=6cm
\centerline{\epsfbox{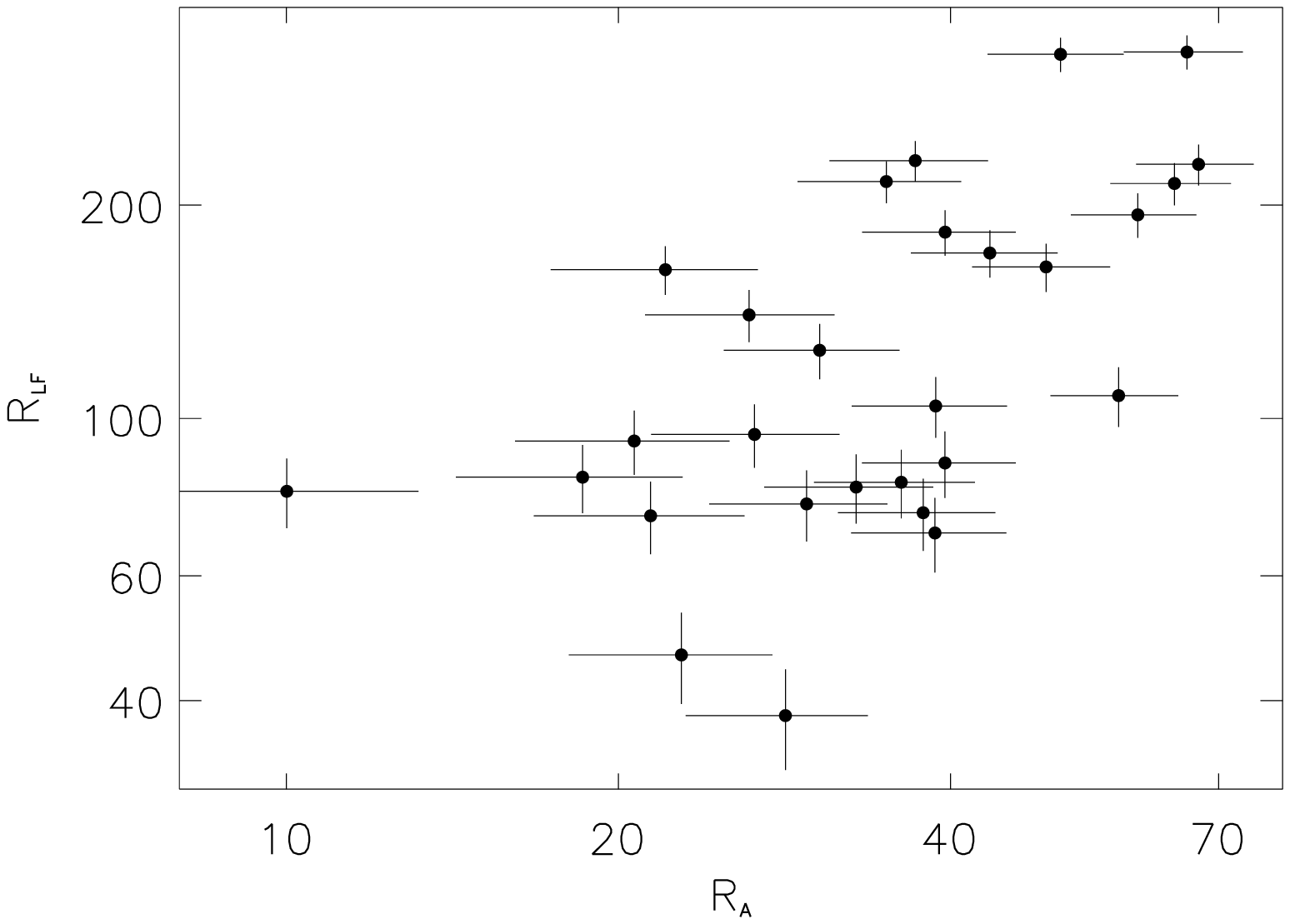}}
 \caption{Comparison of the two richness estimates $r_{A}$ ( based on
 the Abell criterion) and $r_{LF}$ (based on the clusters luminosity
 function - see text for details).\label{fig:rich}}
 \end{figure}

\begin{figure} 
\epsfxsize=10cm
\centerline{\epsfbox{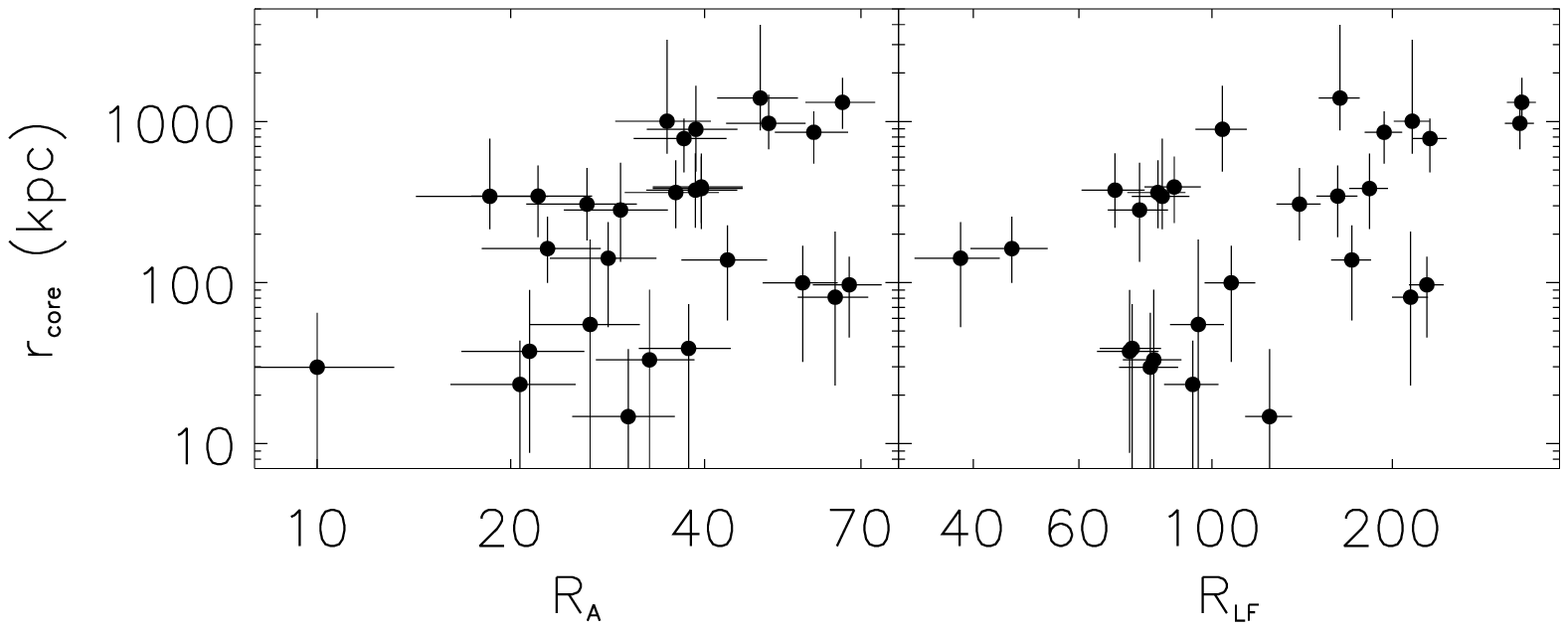}}
 \caption{Core radius versus cluster richness, for both the Abell and
 luminosity function richness estimates. \label{fig:r0rich}}
 \end{figure}

\begin{figure}
\epsfxsize=10cm
\centerline{\epsfbox{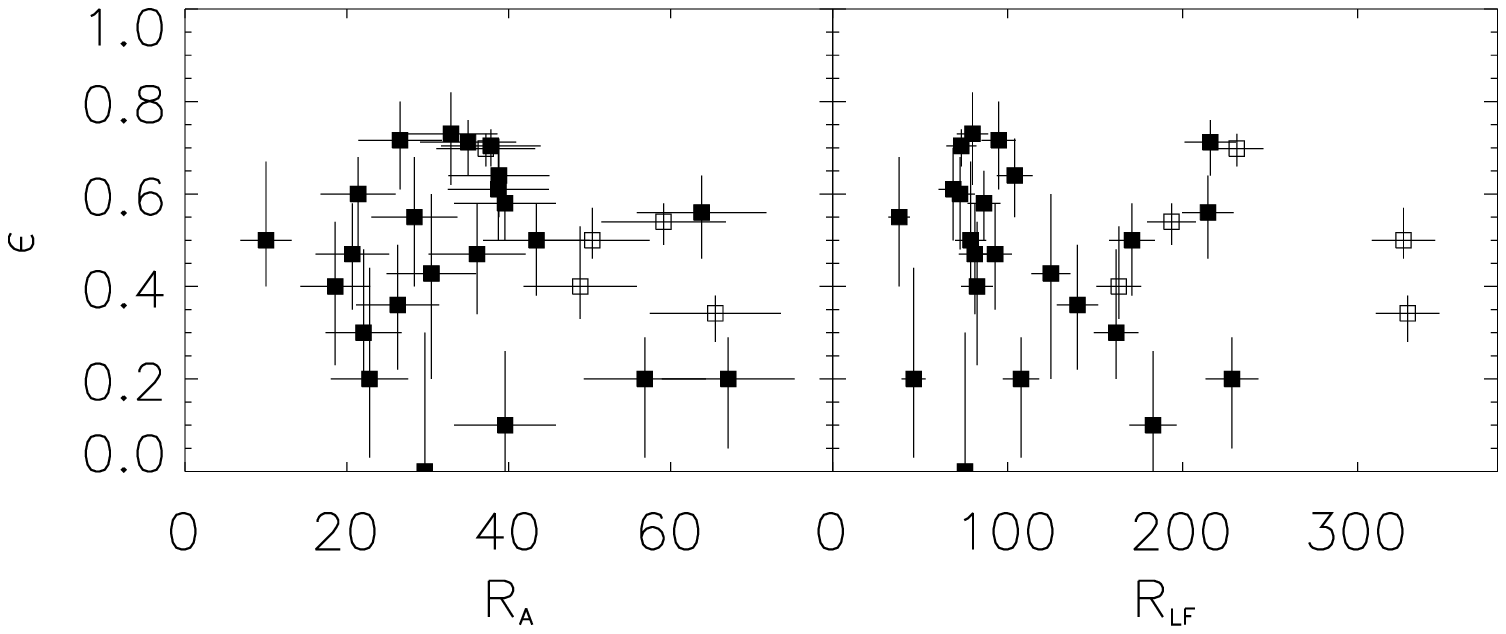}}
\caption{Ellipticity vs. cluster richness, for both the Abell and
luminosity function richness estimates. Empty symbols mark the less
regular clusters (see text for details).  \label{fig:ellrich}}
 \end{figure}

\begin{figure}
\epsfxsize=10cm
\centerline{\epsfbox{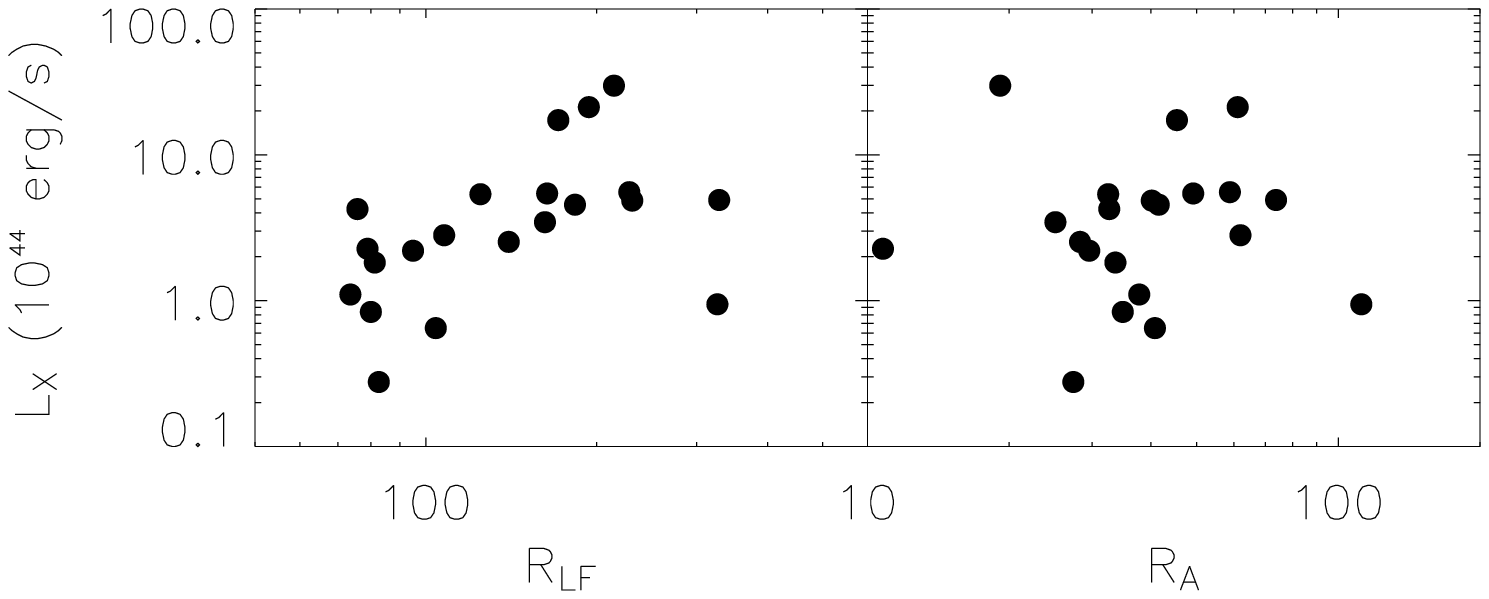}}
\caption{Cluster X-ray luminosity vs. richness, for both the Abell and
luminosity function richness estimates. Only 21 clusters with X-ray luminosities available
are shown .  \label{fig:richlflx}}
\end{figure}

\end{document}